\documentclass[paper=A4,11pt,abstract=true,numbers=noenddot,
titlepage=false,twocolumn=false,
draft=false]{scrartcl}

\makeatletter
\DeclareOldFontCommand{\rm}{\normalfont\rmfamily}{\mathrm}
\DeclareOldFontCommand{\sf}{\normalfont\sffamily}{\mathsf}
\DeclareOldFontCommand{\tt}{\normalfont\ttfamily}{\mathtt}
\DeclareOldFontCommand{\bf}{\normalfont\bfseries}{\mathbf}
\DeclareOldFontCommand{\it}{\normalfont\itshape}{\mathit}
\DeclareOldFontCommand{\sl}{\normalfont\slshape}{\@nomath\sl}

\usepackage[usenames,dvipsnames]{color}
  \definecolor{hgreen}{rgb}{0,.3,0}
  \definecolor{hred}{rgb}{.3,0,0}
  \definecolor{hblue}{rgb}{0,0,.3}
  \definecolor{LightGray}{gray}{0.95}
\usepackage[
	    colorlinks=true,
	    linkcolor=hblue,
	    citecolor=hgreen,
	    filecolor=hblue,
	    urlcolor=hred
	    ]{hyperref}
\usepackage{pdfpages}
\usepackage{graphicx}
\usepackage{mathrsfs}
\usepackage{amsmath}
\usepackage{array}
\usepackage{amssymb}
\usepackage{braket}
\usepackage{bbold}
\usepackage{slashed}
\usepackage{cleveref}
\usepackage{subfig}
\usepackage[titletoc,title]{appendix}
\usepackage[affil-it ]{authblk}
\usepackage{libertine}
\usepackage[numbers,sort&compress]{natbib}

\newcommand{\UV}{{\rm UV}}
\newcommand{\IR}{{\rm IR}}
\newcommand{\MS}{\overline{\rm MS}}
\newcommand{\BP}[3]{\langle #1\rangle^{(#2)}\bigl\vert_{#3}}
\newcommand{\Lag}{\mathscr{L}}
\newcommand{\Q}{\mathscr{Q}}
\newcommand{\E}{\mathscr{E}}

\newcommand{\Z}{\mathcal{Z}}
\newcommand{\cO}{\mathscr{O}}

\newcommand{\oZ}[2]{\Z^{(#1)}_{#2}}

\newcommand{\Nf}{N_{\!f}}
\newcommand{\order}[1]{ {\cal O}(#1)}
\newcommand{\ntr}{ {n_{\rm tr}}}

\newcommand{\tZ}[2]{\tilde{\Z}^{(#1)}_{#2}}

\setcapindent{1em}
\setkomafont{captionlabel}{\bf}
\setkomafont{caption}{\it}
\numberwithin{equation}{section}

\begin{document}

\renewcommand\Authands{, }
\unitlength = 1mm

\title{Operator mixing in $\boldsymbol{\epsilon}$-expansion\\{\LARGE Scheme and evanescent (in)dependence}} 
\date{\today}
\author[a]{Lorenzo Di Pietro%
        \thanks{\texttt{ldipietro@perimeterinstitute.ca}}}
\author[b]{Emmanuel Stamou%
        \thanks{\texttt{estamou@uchicago.edu}}}
\affil[a]{Perimeter Institute for Theoretical Physics, Waterloo, ON\,N2L\,2Y5, Canada}
\affil[b]{Enrico Fermi Institute, University of Chicago, Chicago, IL 60637, USA}

\maketitle

\begin{abstract}

We consider theories with fermionic degrees 
of freedom that have a fixed point of Wilson--Fisher type
in non-integer dimension $d = 4-2\epsilon$. 
Due to the presence of {\it evanescent} operators, i.e., operators that 
vanish in integer dimensions,
these theories contain families of infinitely many operators
that can mix with each other under renormalization.
We clarify the dependence of the corresponding 
anomalous-dimension matrix on the choice of renormalization
scheme beyond leading order in $\epsilon$-expansion.
In standard choices of scheme, we find that 
eigenvalues at the fixed point cannot be
extracted from a finite-dimensional block.
We illustrate
in examples a truncation approach to 
compute the eigenvalues.
These are observable scaling dimensions, and, indeed,
we find that the dependence on 
the choice of scheme cancels.
As an application, we obtain the IR scaling dimension of 
four-fermion operators in QED 
in $d=4-2\epsilon$ at order $\order{\epsilon^2}$.
\end{abstract}

\pdfbookmark[1]{Table of Contents}{tableofcontents}
\setcounter{page}{1}
\tableofcontents

\section{Introduction}
One of the tools to study conformal field theories  (CFTs) is to realize 
them as the endpoint of a renormalization group (RG) flow.
Starting from a description in terms of a weakly-coupled UV Lagrangian 
deformed by a relevant coupling, the dimension $d$ of space(-time) 
can be continued close to the upper-critical value $d_c$, 
in which the IR and the free UV fixed points coalesce.
When $d = d_c - 2 \epsilon$ with $\epsilon \ll 1$,
the observables of the IR CFT admit a systematic expansion 
in the parameter $\epsilon$ \cite{Wilson:1971dc, Wilson:1972cf}. 
Eventually, an extrapolation to $\epsilon$ of $\order{1}$ is attempted
in order to estimate observables of the original strongly-coupled CFT.\footnote{
Even though the fixed point in non-integer dimension
is defined using perturbation theory, 
it is an implicit assumption of the extrapolation
procedure that it exists also beyond the perturbative regime. 
In this paper we only 
consider perturbation theory. For non-perturbative
studies of CFTs in non-integer dimensions, see refs. \cite{El-Showk:2013nia, Chester:2014gqa, Bobev:2015jxa, Chester:2015qca, Chester:2015lej}.}
   
A known property of the dimensional continuation is that the spectrum 
of operators is enlarged, i.e., in $d= d_c -2 \epsilon$ 
there exist so-called {\it evanescent} operators  
that become redundant when $\epsilon \to 0$. 
These operators are more than a mere curiosity: 
as a consequence of their existence, 
the fixed points in non-integer $d$ 
have qualitatively new features compared to the standard, 
integer-dimensional CFTs that they continue. 
For instance, it was shown in refs.~\cite{Hogervorst:2014rta, Hogervorst:2015akt} 
that in theories of free bosons and $\phi^4$-theories 
evanescent operators 
lead to negative-norm states in radial quantization, 
implying that the fixed point in non-integer dimension 
is not unitary. 
These negative norm states have large scaling dimensions,
so in the example considered in ref.~\cite{Hogervorst:2014rta, Hogervorst:2015akt} 
the evanescent operators do not affect the properties of the light spectrum. 

The departure from standard CFTs is even more pronounced
in theories with fermionic degrees of freedom, due to the fact that 
theories with free fermions in non-integer dimension contain 
infinitely many evanescent operators with the same scaling 
dimension and spin. 
One way to construct them is by antisymmetrizing 
$n$ gamma matrices, where $n$ runs over the positive integers, such that they 
vanish in integer $d < n$.  
The simplest example is that of the scalar four-fermion operators 
\begin{equation}\label{eq:evaexample}
\cO_{n} = (\overline{\Psi} \Gamma^{n}_{\mu_1\dots \mu_n}\Psi )^2~,
\quad\text{with}\quad \Gamma^{n}_{\mu_1\dots \mu_n}\equiv \gamma_{[\mu_1} \dots \gamma_{\mu_n]}~,
\end{equation}
where the square brackets denote antisymmetrization. 
When interactions are turned on, all these operators can mix with each other.
These mixings result in an anomalous-dimension matrix (ADM) of infinite size, 
which makes the computation of the eigenvalues a considerably more 
involved problem than in the bosonic case.

At leading order (LO), there is a drastic simplification, 
because the operators that are not evanescent ---the so-called {\it physical} operators\footnote{
We borrow the 
nomenclature from the literature on dimensional regularization, in which the
physical theory lives in integer dimension. Note, however, that in fractional 
dimensions the evanescent operators are as physical as the non-evanescent 
ones.}--- 
form a finite-dimensional invariant subspace under mixing, i.e., 
the evanescent--physical entries of the LO ADM vanish. 
Therefore, the IR scaling dimensions of the physical operators 
at LO in $\epsilon$ can be obtained by diagonalizing 
a finite-size matrix.
At next-to-leading order (NLO) and beyond, the ADM is scheme-dependent. 
In the context of $d=4$ computations within dimensional regularization, 
refs.~\cite{Buras:1989xd,Dugan:1990df} 
introduced a scheme choice with the attractive property that
the block form of the LO ADM is preserved at all orders.
The same scheme was proposed in refs.~\cite{Bondi:1988fp,Bondi:1989nq} in the context of 
$d=2$ Gross--Neveu/Thirring models, though formulated 
in a different language.

In this paper we investigate the problem of obtaining the 
IR scaling dimension of physical operators beyond LO in $\epsilon$ 
when there is mixing with infinitely many evanescent operators.
Naively, the scheme choice of ref.~\cite{Buras:1989xd,Dugan:1990df,Bondi:1988fp,Bondi:1989nq} seems to 
trivialize the problem by restricting it to the finite-dimensional
invariant subspace spanned by physical operators.
However, this leads to an apparent paradox,
because the entries of the finite-size block of the ADM do 
not transform properly under a change of basis. 
For instance, they are affected by redefinitions of the evanescent 
operators by a multiple of $\epsilon \times$the physical 
operators~\cite{Herrlich:1994kh}. 
This can be interpreted as a sign of 
renormalization-scheme dependence.

We resolve this issue by studying the transformation rules of the ADM
under a change of scheme, keeping $\epsilon \neq 0$. 
The general transformation rule turns out to depend on $\epsilon$. 
We demonstrate how this $\epsilon$ dependence is important 
to ensure that the eigenvalues at the fixed point are invariant under 
a change of scheme. 
In particular, we show that going from the minimal subtraction ($\MS$) 
scheme to the scheme used for evanescent operators in 
ref.~\cite{Buras:1989xd,Dugan:1990df,Bondi:1988fp,Bondi:1989nq} introduces terms of order $\epsilon$ in the one-loop 
ADM.\footnote{We are here assuming to be in the generic case 
in which the ADM contains 
physical--evanescent entries already at one-loop. 
In this case, the evanescent operators affect the anomalous dimension starting at $\order{\epsilon^2}$.
More generally, the physical--evanescent mixing may first start at order $L$
in perturbation theory, in which case the evanescent operators affect the 
anomalous dimension starting at $\order{\epsilon^{L+1}}$.
For instance, in the Gross--Neveu model $L=3$ \cite{Gracey:2016mio}.} 
At the fixed point, these terms spoil the block structure 
of the ADM at $\order{\epsilon^2}$. 
Therefore, the finite-size physical block of the two-loop ADM is 
not sufficient  by itself to extract the scaling dimensions at NLO.
The additional input required is the full one-loop mixing of the
infinite tower of evanescent operators into the physical operators.
Once this is known, one can finally compute the finitely many 
eigenvalues associated  to the physical operators.
This requires a rotation or, equivalently, a further change of scheme that 
completely fixes the aforementioned ambiguity in the choice of 
basis for the evanescent operators.  
We illustrate the procedure by carrying it out explicitly in the 
example of four-fermion operators in QED in $d=4-2\epsilon$. 
After computing all the relevant entries of the ADM up to order 
$\epsilon^2$, we approximate the first two eigenvalues by 
truncating to a finite number of evanescent operators.
We test that the approximations converge
as we increase the truncation.

There are several examples of CFTs with fermionic degrees of freedom that can be 
studied in $\epsilon$-expansion and to which the method we describe here 
is applicable, see for instance the recent works 
\cite{DiPietro:2015taa, Giombi:2015haa, Fei:2016sgs, 
Gracey:2016mio, Diab:2016spb, Giombi:2016fct, Giombi:2017rhm, Dalidovich:2013qta, Sur:2016nhs, PhysRevLett.102.046406, Lunts:2017lun, Janssen:2017eeu} and references therein. 
In our companion paper \cite{DiPietroStamou2}, we focus on $3d$ QED and 
use the NLO eigenvalues obtained here to estimate the scaling dimensions
of four-fermion operators in $d=3$.

The rest of the paper is organized as follows: 
in section~\ref{sec:EpsRev} we review 
the general setup of the $\epsilon$-expansion, fix our notation, and relate the CFT scaling 
dimensions at NLO to renormalization constants; 
in section~\ref{sec:SchDep} we discuss the transformation rules of the beta 
function and the ADM under a change of 
renormalization scheme, first to all orders in perturbation 
theory and then more explicitly at the two-loop order, 
illustrating the scheme-independence of the scaling dimensions; 
in section~\ref{sec:EvaOpe} we explain the block structure of the 
mixing between evanescent and physical operators and show that neither in $\MS$ 
nor in the scheme of ref.~\cite{Buras:1989xd,Dugan:1990df,Bondi:1988fp,Bondi:1989nq} does the ADM at the fixed point 
have an invariant subspace at NLO; 
in section~\ref{sec:EvaTow} we work out the example of four-fermion 
operators in QED and introduce the truncation algorithm that allows us to compute 
the scaling dimensions at the fixed point.
Supplementary material and formulas for the anomalous dimensions 
computed in ref.~\cite{DiPietroStamou2} are collected in the appendices.
 
\section{Fixed points and scaling dimensions in $\boldsymbol{d=4-2\epsilon}$\label{sec:EpsRev}}

Consider a theory in $d=4$ that admits a perturbative expansion 
in a classically marginal coupling $\alpha$. For $\alpha = 0$ the theory 
is free; its local operators are products of the fields and their 
derivatives, and their correlators are given by Wick contractions. 
When the interaction is turned on, we can compute corrections to the 
correlators in a perturbative expansion in $\alpha$. Each order in 
this expansion can be continued to a non-integer value of the dimension 
$d$ \cite{Wilson:1972cf, Collins:1984xc}. Upon continuation to $d \neq 4$, 
the coupling acquires a nonzero mass dimension. For definiteness 
we keep in mind the example of gauge theories, where 
$\alpha = \frac{g^2}{16\pi^2}$, and take this acquired 
mass dimension to be $4-d\equiv 2 \epsilon$. 

Correlators of local operators have poles at $\epsilon = 0$, which can be subtracted by 
defining the renormalized coupling and the renormalized operators 
as
\begin{align}
\alpha_0 & = Z_\alpha \alpha(\mu) \mu^{2 \epsilon} \label{eq:rencoupling} ~,\\
(\cO_0)_i & = (\Z^{-1})_i^{~j} \cO_j  \label{eq:renoperator} ~.
\end{align}
The subscript ``0'' labels bare quantities. 
 $Z_\alpha$ 
and $(\Z^{-1})_i^{~j}$ are the renormalization constants that
subtract the divergences.\footnote{In equation \eqref{eq:renoperator} both the bare and the renormalized operator are thought of as products of bare elementary fields. The constants $\Z$ subtracts only the divergences that arise from the products of fields at the same point and do not include the wave-function renormalizations of the elementary fields.}
We stress that here we are interested in the dynamics
for $\epsilon \neq 0$. Therefore, the procedure of absorbing the 
divergences in $Z_\alpha$ 
and $(\Z^{-1})_i^{~j}$ is just  an efficient way to keep track of the leading behavior
of correlators for $\epsilon \ll 1$. This observation also appeared 
recently in ref. \cite{Behan:2017emf}, see also section $1.35$ of ref. \cite{Vasilev:2004yr}.

In the perturbative expansion of the renormalization
constants in $\alpha$, each term
admits an additional Laurent expansion in $\epsilon$, i.e.,
\begin{align}
Z_\alpha(\alpha, \epsilon) = 1 + \sum_{L=1}^{\infty}\alpha^L \sum_{M = - L}^{\infty} \epsilon^{M} Z_\alpha^{(L,-M)}~, \label{eq:expconst1}\\
\Z(\alpha, \epsilon) = \mathbb{1} + \sum_{L=1}^{\infty}\alpha^L \sum_{M=-L}^{\infty} \epsilon^{M} \oZ{L,-M}{} ~. \label{eq:expconst2}
\end{align} 
Different choices of the terms that are finite for $\epsilon \to 0$ 
correspond to different (mass-independent) renormalization schemes.
A standard choice is the $\MS$ scheme,\footnote{
We do not distinguish between $\rm{MS}$ and $\MS$ or any ${\rm MS}$-like 
scheme, in which the same constant term is always subtracted together 
with a pole  \cite{Bardeen:1978yd}. 
In the generic mass-independent schemes we consider, the finite  terms in the renormalization constants are totally
arbitrary, i.e., they cannot be reabsorbed with an overall rescaling of $\mu$ as in ${\rm MS}$-like 
schemes.
}
in which these finite terms are set to zero. 
When evanescent operators are present, it is more convenient 
to use a different scheme  that includes some specific finite terms $\oZ{L,0}{}$ 
We discuss this in detail below. For the moment, we keep the scheme generic. 

From the renormalization constants one obtains the RG functions, namely the beta
function and the ADM. 
The beta function determines the running 
of the coupling $\alpha$ 
\begin{equation}\label{eq:betadef}
\frac{d \alpha }{d\log\mu} = -2 \epsilon \alpha +  \beta(\alpha, \epsilon)~.
\end{equation}
A convenient way to define the ADM is to consider the theory deformed 
by adding new couplings proportional to composite operators
\begin{equation}\label{eq:Lagdeform}
\Lag \to \Lag + (C_0)^i (\cO_0)_i ~.
\end{equation}
The ADM, $\gamma$, is defined as the running of the 
couplings $C^i$ to linear order in them, namely
\begin{equation}\label{eq:gammadef}
{\gamma(\alpha, \epsilon)}_j^{~i}\equiv
\frac{\partial}{\partial C^j}\left.\left(\frac{d C^i}{d\log\mu} \right)\right\vert_{C = 0}\,. 
\end{equation}
It then follows from eq.~\eqref{eq:renoperator} that the renormalized 
couplings $C^i$ are related to the bare ones via
\begin{equation}\label{eq:renC}
(C_0)^j = C^i  \Z_i^{~j}~.
\end{equation}
From the fact that  
bare quantities do not depend on the renormalization scale $\mu$,
we obtain via eqs.~\eqref{eq:betadef} and \eqref{eq:gammadef}
the standard  formulas
\begin{align}
\beta(\alpha, \epsilon)& =  
-\alpha \frac{d \,\log Z_\alpha}{d \log\mu} \equiv 
-2\alpha \sum_{L=1}^\infty \alpha^L \beta^{(L)} ~, \label{eq:expbeta} \\ 
{\gamma(\alpha, \epsilon)} & = - \frac{d \log \Z }{d \log\mu} \equiv  
\sum_{L=1}^\infty\alpha^L \gamma^{(L)} ~.\label{eq:expgamma}
\end{align}
In schemes in which finite terms $Z_\alpha^{(L,M)}$ and  $\oZ{L,M}{}$ 
with $M\leq0$ are present in the renormalization constants, 
$\beta$ and $\gamma$ contain terms with positive powers of $\epsilon$.
We shall keep track of them in order to discuss the scheme independence 
of observables in the next section. 
We define them via the expansions
\begin{align}
\beta^{(L)} & = \sum_{M= 0}^{\infty }\beta^{(L,-M)}\epsilon^M ~, \label{eq:oepsbeta} \\
\gamma^{(L)} & = \sum_{M= 0}^{\infty }\gamma^{(L,-M)}\epsilon^M ~. \label{eq:oepsgamma}
\end{align}

We are interested in studying non-trivial fixed points of the RG 
in $d=4-2\epsilon$ with $\epsilon > 0$. These are defined by the condition 
\begin{equation}\label{eq:fixpoint}
\left. \frac{d \alpha}{d \log \mu}\right\vert_{\alpha^*} = -2 \epsilon \alpha^* + \beta(\alpha^*, \epsilon) 
= 0~.
\end{equation}
The solution of the above condition for $\epsilon \ll 1$, 
up to second order in $\epsilon$ is
\begin{equation}
\alpha^*=-\epsilon   \frac{1}{\beta^{(1,0)}} 
         -\epsilon^2 \frac{\beta^{(2,0)}-\beta^{(1,0)}\beta^{(1,-1)}}{ {\beta^{(1,0)}}^3}+\order{\epsilon^3}~.
\label{eq:astar}
\end{equation}
By requiring $\alpha^* > 0$, we find that an IR fixed point exists 
only if $\beta^{(1,0)} < 0$, i.e., when the coupling is marginally irrelevant 
in $d=4$.
(This is, of course, because we assumed the mass dimension of 
$\alpha$ to be positive in $d < 4$; alternatively, one could have a 
marginally relevant coupling, which acquires a negative mass dimension, 
and find a perturbative UV fixed point.)

In the free UV theory, the scaling dimensions of operators are just 
the sum of the canonical dimensions of the free fields that compose 
them; we denote these UV scaling dimensions by $\Delta_{\UV}$. 
The ADM has a block form, in the sense that  only 
operators with the same spin and the same value of $\Delta_{\UV}$ can mix.
Within each block, the scaling dimensions of operators at the IR 
fixed point are given by
the eigenvalues of 
\begin{equation}
\Delta_{\UV} \mathbb{1} + \gamma_*~,
\end{equation}
where $\gamma_*$ is the ADM evaluated at 
the fixed point.
This can be derived by applying the RG equation to the two-point 
correlation function \cite{Mitter:1973ue, Brezin:1973jc, Brezin:1974eb}. 
See appendix~\ref{app:Resummation} for a derivation. 
Up to second order in $\epsilon$ we have that
\begin{equation}\label{eq:gammastar}
\gamma^*  \equiv \gamma(\alpha^*, \epsilon) = \epsilon \gamma_1^* +\epsilon^2\gamma_2^* + \order{\epsilon^3}~,   
\end{equation}
where
\begin{align}
\gamma_1^* & = -\frac{\gamma^{(1,0)}}{\beta^{(1,0)}}~, \label{eq:gamma1star}\\
\gamma_2^* & = \frac{
  \beta^{(1,0)} \gamma^{(2,0)} 
- \beta^{(2,0)} \gamma^{(1,0)}
}{{\beta^{(1,0)}}^3}+
\frac{
   \beta^{(1,-1)} \gamma^{(1,0)}
- {\beta^{(1,0)}} \gamma^{(1,-1)}
}{{\beta^{(1,0)}}^2}~.\label{eq:gamma2star}
\end{align}
All terms on the right hand side of eqs.~\eqref{eq:gamma1star} and 
\eqref{eq:gamma2star} are fixed in terms of renormalization constants.
We collect these relations in appendix~\ref{app:BetaAD}.
The relevant aspect is that, while $\gamma_1^*$ does not depend on finite
renormalization constants, i.e., it is scheme independent, both 
terms of $\gamma_2^*$ do depend on such finite constants.

$\Delta_{\UV}$ is just a constant shift within each block, so to obtain 
the IR scaling dimension up to $\order{\epsilon^2}$ it is sufficient
to perturbatively diagonalize $\frac{1}{\epsilon}\gamma_* = \gamma_1^* + 
\epsilon \gamma_2^* + \order{\epsilon^2}$. 
At this order, the corresponding set of eigenvalues are
\begin{equation}
(\gamma_1^*)_i + \epsilon (U \gamma_2^*U^{-1})_i^{~i} + \order{\epsilon^2}~,
\end{equation} 
where $(\gamma_1^*)_i$ denotes the $i$-th eigenvalue of $\gamma_1^*$, 
$U$ is the rotation to the basis of eigenvectors of $\gamma_1^*$, 
i.e., $U \gamma_1^* U^{-1} = \text{diag}[(\gamma_1^*)_i]$, and 
$(U \gamma_2^*U^{-1})_i^{~i}$ is the $i$-th diagonal matrix element 
of $\gamma_2^*$ in the rotated basis. The IR scaling dimension of 
the $i$-th operator with UV dimension $\Delta_{\UV}$ then equals
\begin{align}\label{eq:IRscaldimpert}
(\Delta_{\IR})_i & = \Delta_{\UV} + \epsilon (\Delta_1)_i + \epsilon^2 (\Delta_2)_i  + \order{\epsilon^3}\,,
\end{align}
with the definitions
\begin{align}
\label{eq:D1andD2}
(\Delta_1)_i & \equiv (\gamma_1^*)_i\qquad\text{and}\qquad (\Delta_2)_i \equiv (U \gamma_2^*U^{-1})_i^{~i}~.
\end{align}

\section{Scheme independence of scaling dimensions \label{sec:SchDep}}

The scaling dimensions are observables. Therefore, they cannot depend on the 
subtraction scheme that we use to compute the renormalization constants. 
This is not evident from eqs.~\eqref{eq:IRscaldimpert} and \eqref{eq:D1andD2}, 
because both the ADM and the beta function, which 
are used to define $\gamma_*$, do depend on the scheme. We now explain 
how the scheme dependence cancels in the eigenvalues. 
Even though this is a well-known result 
(see for instance section~1.40 of ref.~\cite{Vasilev:2004yr}), it is useful for 
us to review it here, because we shall make use of it later to identify 
a convenient scheme for the case in which evanescent operators are present. 
We first review the general 
argument at all orders in perturbation theory and then present 
the explicit formulas for the change of scheme up to two-loop order.

Renormalization schemes are parametrized by the coefficients of the 
finite terms, $Z_\alpha^{(L,M)}$ and  $\oZ{L,M}{}$ with $M\leq0$. 
We denote the renormalization constants of a new scheme by $\tilde{Z}_\alpha$ 
and $\tilde{\Z}$.
The definitions in eqs.~\eqref{eq:rencoupling} and \eqref{eq:renC} imply that 
\begin{align}
\tilde{\alpha}\,\tilde{Z}_\alpha(\tilde{\alpha},\epsilon) & = \alpha \,Z_\alpha(\alpha,\epsilon)~,\\
\tilde{C}^i\tilde{\Z}_i^{~j}(\tilde{\alpha},\epsilon) & = C^i\Z_i^{~j}(\alpha,\epsilon)~.
\end{align}
The first line defines the renormalized coupling in the new scheme, 
$\tilde{\alpha}$, as a function of $\alpha$ and $\epsilon$. 
Since the divergent terms agree, i.e., $Z_\alpha^{(L,M)} = \tilde{Z}_\alpha^{(L,M)}$ 
for $M>0$, the Laurent expansion of 
$\tilde{\alpha} = \tilde{\alpha}(\alpha, \epsilon)$ cannot contain 
negative powers of $\epsilon$. 
We can then define the change of scheme to all orders in perturbation theory
via functions that depend solely on $\epsilon$ and $\alpha$, i.e.,
\begin{equation}\label{eq:changsch1}
\tilde{\alpha} = f(\alpha, \epsilon) \alpha
\quad\text{and}\quad
\tilde{C}^i =  C^j F_j^{~i}(\alpha,\epsilon)\,,
\end{equation}
where
\begin{equation}
\label{eq:changsch2F}
F(\alpha, \epsilon) \equiv \Z(\alpha,\epsilon)\tilde{\Z}^{-1}(\tilde{\alpha}(\alpha),\epsilon)\,,
\end{equation} 
with the normalizations $f(0,\epsilon) = 1$, and $F(0,\epsilon) = \mathbb{1}$ and 
both functions regular at $\epsilon= 0$. 

Since
\begin{equation}\label{eq:chainbeta}
\frac{d \tilde{\alpha} }{d\log\mu} = \partial_\alpha(f(\alpha,\epsilon) \alpha ) \frac{d \alpha }{d\log\mu}~, 
\end{equation}
the fixed point in $\alpha$ gets mapped to the fixed point in 
$\tilde{\alpha}$, i.e., $\tilde{\alpha}^* = f(\alpha^*,\epsilon) \alpha^*$. 

As for the anomalous dimension, we have that
\begin{align}\label{eq:schgamma}
\tilde{\gamma}_j^{~i} & = \left.\left((F^{-1})_j^{~k} \frac{\partial}{\partial C^k}   \frac{d C^l}{d \log \mu} F_l^{~i}\right)\right\vert_{C=0} + (F^{-1})_j^{~k} \partial_\alpha F_k^{~i} \frac{d \alpha }{d\log\mu}\nonumber\\ & = \left((F^{-1})_j^{~k} \gamma_k^{~l} F_l^{~i}\right) + (F^{-1})_j^{~k} \partial_\alpha F_k^{~i} \frac{d \alpha }{d\log\mu}~.
\end{align}
In the evaluation of eq.~\eqref{eq:schgamma} at $\tilde{\alpha}^*$, the second 
term drops out, and we see that the matrix at the fixed point is affected 
by the change of scheme only through a similarity transformation with the matrix $F$.
This does not affect the eigenvalues, thus proving
that the scaling dimensions are scheme-independent.

Next, we show which terms enter the cancellation of the scheme dependence 
in perturbation theory, up to two-loop order.
To this end, we must first relate the one- and two-loop coefficients of the 
beta function and ADM in the two schemes; we list these relations 
in appendix~\ref{app:BetaAD}.
Using them, we evaluate the anomalous dimensions at the fixed point 
via eqs.~\eqref{eq:gamma1star} and \eqref{eq:gamma2star} to obtain
\begin{align}
\tilde{\gamma}_1^*&= \gamma_1^*~, \label{eq:schgamma1star}\\
\tilde{\gamma}_2^*&= \gamma_2^*  
                   \underbrace{-\frac{1}{\beta^{(1,0)}} \left[\gamma_1^*, \oZ{1,0}{}- \tZ{1,0}{}\right]}_{
		   \equiv \delta\gamma_2^*}~.
\label{eq:schgamma2star}
\end{align}
These equations can be understood as the perturbative expansion of 
the result in eq.~\eqref{eq:schgamma} evaluated at the fixed point. 
Since the difference, $\delta \gamma_2^*$, is  a commutator 
with $\gamma_1^*$, one readily derives that $(U \delta \gamma_2^* U^{-1})_i^{~i} = 0$,
which means that the IR scaling dimension of eq.~\eqref{eq:IRscaldimpert} 
does not change with the scheme shift.
This is the NLO manifestation of the scheme 
independence of the scaling dimensions.

We stress that the NLO scheme independence of the scaling dimension requires 
that we include the term $\gamma^{(1,-1)}$ in $\gamma_2^*$, see 
eq.~\eqref{eq:gamma2star}. $\gamma^{(1,-1)}$ is the coefficient of the term linear in $\epsilon$ 
in the one-loop anomalous dimension and depends on the scheme 
as shown in eq.~\eqref{eq:pertchangegamma}. 
Clearly such a $\order{\epsilon}$ term would be disregarded in 
the computation of the 
anomalous dimension in $d=4$, but we must retain 
it when we compute observables at the fixed point in $d=4-2\epsilon$. 
More generally, this applies at higher orders to all the terms with 
positive powers of $\epsilon$ that appear in the beta function 
and the anomalous dimensions in a generic scheme.

\section{Evanescent operators \label{sec:EvaOpe}}

In the free theory at $\alpha = 0$ local operators can be defined 
by (gauge-invariant) products of the free fields and their derivatives. 
These composite operators often satisfy linear relations
that reduce the number of independent monomials in the fields. 
However, many relations (in fact, infinitely many) are satisfied
when $\epsilon = 0$ but are violated by positive powers of $\epsilon$. 
More generally, many relations hold only if $d$ is integer. 
This implies that in non-integer dimension 
there are additional independent operators, which are 
called {\it evanescent} operators 
because they vanish when $\epsilon \to 0$.

For instance, any operator defined through the antisymmetrization 
of $n$ indices, such as the four-fermion operator
\begin{equation}\label{eq:evaexample2}
\cO_{n} = (\overline{\Psi} \Gamma^{n}_{\mu_1\dots \mu_n}\Psi )^2~,
\quad\text{with}\quad \Gamma^{n}_{\mu_1\dots \mu_n}\equiv \gamma_{[\mu_1} \dots \gamma_{\mu_n]}~,
\end{equation}
is equal to zero
for all integer values of $n > d$, because 
there are not enough possible values for the indices to antisymmetrize 
$n$ of them. 
(The square brackets denote antisymmetrization normalized as
$\gamma_{[\mu_1} \dots \gamma_{\mu_n]} \equiv \frac{1}{n!}\sum_\sigma (-1)^\sigma \gamma_{\mu_{\sigma(1)}} \dots \gamma_{\mu_{\sigma(n)}} $.)
However, $\cO_n$ is a non-trivial operator when $d$ is non-integer, 
as can be seen by considering the associated Feynman rule, i.e., the 
contraction with two $\Psi$ and two $\overline{\Psi}$ fields, which reads
\begin{equation}\label{eq:structureEn}
S^n_{\alpha\beta\gamma\delta} \propto (\Gamma^{n\,\mu_1\dots \mu_n})_{\alpha\beta} (\Gamma^{n}_{\mu_1\dots \mu_n})_{\gamma \delta}~.
\end{equation} 
Using the standard rules for the Clifford algebra in $d$ dimensions
\begin{equation}
\{\gamma_\mu, \gamma_\nu\} = 2 \eta_{\mu\nu} \mathbb{1}\,,
\end{equation}
with $\delta^\mu_\mu = d$, we obtain that \cite{Kennedy:1981kp}
\begin{equation}
S^n_{\alpha \beta \beta \alpha} \propto \mathrm{Tr}[\Gamma^{n\,\mu_1\dots \mu_n}\Gamma^{n}_{\mu_1\dots \mu_n}] = \mathrm{Tr}[\mathbb{1}] (-1)^\frac{n(n-1)}{2}\frac{\Gamma(d +1)}{\Gamma(d +1 - n)}~,
\end{equation}
which is non-zero except if $d$ is an integer smaller than $n$.
This demonstrates that, in general, also the structure in eq.~\eqref{eq:structureEn} 
is non-trivial. 
For our purposes we consider an expansion around $d=4$ 
and thus we call evanescent the operators 
in $d = 4 - 2 \epsilon$  that vanish when $\epsilon \to 0$, 
such as $\cO_n$ for $n>5$.
Similarly, one can define evanescent operators relative to 
other integer values of $d$\,.

When interactions are turned on, physical 
and evanescent operators can mix. 
In fact, evanescent operators were first introduced for the computation
of anomalous dimensions in $d = 4$ in dimensional regularization 
in refs.~\cite{Buras:1989xd, Dugan:1990df}, 
because this mixing affects the result for the physical operators. 
(Here, by mixing between operators we mean a corresponding non-zero entry 
in the renormalization constant $\Z$. In particular, the expression 
``$\cO_i$ mixes into $\cO_j$'' means that $\Z_i^{~j} \neq 0$.) 

Due to this mixing, as we flow to the IR fixed point in $d<4$, the 
eigenoperators, i.e., operators with definite scaling dimensions,
become linear combinations  of physical and evanescent operators.
Evanescent operators at the Wilson--Fisher 
fixed point of a scalar field theory in $d<4$ dimensions were recently studied 
in ref.~\cite{Hogervorst:2015akt}, where it was shown that, in general, they 
lead to loss of unitarity if  $d$ is not integer.
There is an important difference between the scalar theory considered in 
ref.~\cite{Hogervorst:2015akt} and a theory with fermions.
In the scalar theory, for any fixed $\Delta_{\UV}$, there is a finite 
number of evanescent operators with this value of $\Delta_{\UV}$.
In the fermionic theory, an infinite number of them may be present, 
as illustrated by the operators $\{\cO_{(n)}\}_{n \in \mathbb{N}}$ 
that all have $\Delta_{\UV} = 2(d-1)$ for $\alpha = 0$.

\subsection{Block structure of the anomalous-dimension matrix}
Even before specifying the interactions, it is possible to draw 
some general conclusions on the form of the mixing between physical 
and evanescent operators. 
Consider a set of 
operators with the same $\UV$
dimension, and 
let us split them into physical and evanescent components, denoted 
collectively by $\Q$ and $\E$, respectively, i.e.,
\begin{equation}
\cO = \begin{bmatrix} \Q \\ \E \end{bmatrix}~.
\end{equation}
We add these operators to the Lagrangian with bare couplings 
$((C_0)_\Q^i, (C_0)_\E^a)$
\begin{equation}
\Lag \to \Lag + \sum_{i} (C_0)_\Q^i \Q_i + \sum_{a} (C_0)_\E^a \E_a
\end{equation}
and compute the mixing matrix $\Z$ by renormalizing these couplings.

To this end, consider the interaction vertices between the 
elementary fields that are proportional to $((C_0)_\Q^i, (C_0)_\E^a)$. 
Each coupling has a particular vertex structure associated 
to it at the tree level
\begin{equation}
V^{(0)} = \sum_{i} (C_0)_\Q^i S_{\Q_i}+ \sum_{a} (C_0)_\E^a S_{\E_a}~.
\end{equation}
The structures $S_{\E_a}$ vanish in the limit $\epsilon \to 0$. 
For instance, in a theory 
of a Dirac fermion, the 
four-fermion operators in eq.~\eqref{eq:evaexample2} 
give rise to the four-fermion vertices
\begin{equation}
\left(V^{(0)}_{\Psi\overline{\Psi}\Psi\overline{\Psi}}\right)_{\alpha\beta\gamma\delta} = \sum_{n = 0}^4 (C_0)^n_{\Q} S^n_{\alpha\beta\gamma\delta}+ \sum_{n=5}^{\infty} (C_0)^n_{\E} S^n_{\alpha\beta\gamma\delta}~,
\end{equation}
where $S^n_{\alpha\beta\gamma\delta} \propto (\Gamma^{n\,\mu_1\dots \mu_n})_{\alpha\beta} (\Gamma^{n}_{\mu_1\dots \mu_n})_{\gamma \delta}$.

Perturbative corrections to the vertex, order by order in the 
coupling $\alpha$, can again be expressed as a linear combination 
of the structures $(S_{\Q_i}, S_{\E_a})$. 
For this step, it is important that $(S_{\Q_i}, S_{\E_a})$ form 
a complete basis of structures. 
The $L$-loop correction to the vertex then is
\begin{align}
V^{(L)} & = \alpha^L \sum_i (C_0)_\Q^i \left(\sum_j  (A^{(L)}_{\Q\Q})_i^{~j} S_{\Q_j} + \sum_b  (A^{(L)}_{\Q\E})_i^{~b} S_{\E_b}\right)
 \nonumber \\ 
& + \alpha^L \sum_a (C_0)_\E^a \left(\sum_j  (A^{(L)}_{\E\Q})_a^{~j} S_{\Q_j} + \sum_b  (A^{(L)}_{\E\E})_a^{~b} S_{\E_b}\right)~,
\end{align}
where the coefficients $A^{(L)}$ contain poles when
$\epsilon\to 0$.\footnote{In ref. \cite{DiPietroStamou2} we use the notation
\begin{equation}
(A^{(L)}_{\cO\cO'}) S_{\cO'} \equiv \BP{\cO}{L}{S_{\cO'}} \nonumber~.
\end{equation} } 
These are subtracted by the renormalization constants $\Z$ that 
define the renormalized couplings via eq.~\eqref{eq:renC}. 
Typically, the $L$-loop coefficients $A^{(L)}$  have a 
leading $\epsilon^{-L}$ pole and also subleading ones. 
However, in the corrections to the evanescent vertices, i.e., the terms 
proportional to $(C_0)^a_\E$,
the projection to the physical 
structures $S_{\Q_j}$ are always accompanied with additional positive 
powers of $\epsilon$ \cite{Dugan:1990df,Herrlich:1994kh}. 
This has important consequences for the 
structure of the matrix $\Z$. 

At one-loop, $A^{(1)}_{\Q\Q}$, $A^{(1)}_{\Q\E}$, and $A^{(1)}_{\E\E}$ 
have $\frac{1}{\epsilon}$ poles, while 
$A^{(1)}_{\E\Q}$ is finite due to the additional factor of 
$\epsilon$ coming from the projection. 
This results in a block form 
for the one-loop renormalization constant $\Z^{(1,1)}$ 
and consequently for the one-loop ADM, 
with zero entries in  the $\E\Q$ block
\begin{equation}
\gamma^{(1,0)} = 2 \oZ{1,1}{}  = 2\begin{bmatrix} \oZ{1,1}{\Q\Q} & \oZ{1,1}{\Q\E} \\ 0 & \oZ{1,1}{\E\E} \end{bmatrix} ~. \label{eq:oneloopblock}
\end{equation}
At two-loop order, the choice of scheme begins to affect the mixing constants and thus the ADM. 
A convenient choice is to use a slightly modified $\MS$ prescription that
subtracts the finite terms in the  $\E\Q$ block \cite{Dugan:1990df}.
In this scheme, the mixing constant $\Z_{\E\Q}$ is chosen to 
cancel the finite term $A^{(1)}_{\E\Q}$. 
The finite, one-loop terms then are
\begin{align}
Z_\alpha^{(1,0)} & = 0~, \\
\oZ{1,0}{} & = \begin{bmatrix} 0 & 0 \\ \oZ{1,0}{\E\Q} & 0 \end{bmatrix}~.\label{eq:oneloopfinite}
\end{align}
The motivation for choosing this scheme is that 
it simplifies the structure of the two-loop 
ADM, as we shall explain next.

At two-loop order, $A^{(2)}_{\Q\Q}$, $A^{(2)}_{\Q\E}$, and $A^{(2)}_{\E\E}$ 
can contain both $\frac{1}{\epsilon^2}$ and $\frac{1}{\epsilon}$ poles. 
The coefficient of the $\frac{1}{\epsilon^2}$ 
divergence is fixed by the one-loop result, i.e.,
the renormalization 
constants satisfy the RG identity
\begin{equation}
\oZ{2,2}{} = \frac 12 \oZ{1,1}{}\oZ{1,1}{} - \frac{1}{2} \beta^{(1,0)}\oZ{1,1}{}~,\,\label{eq:RGidentity}
\end{equation}
which ensures that the ADM is free of divergences. 
The subleading $\frac{1}{\epsilon}$ 
divergences, $\oZ{2,1}{}$, determine the two-loop ADM. 
In the $A^{(2)}_{\E\Q}$ block, the divergences are still
down by a factor of $\epsilon$ due to the projection, but now this 
does not mean that the mixing constant is finite, because 
a $\frac{1}{\epsilon^2}$ divergence from a loop integral
can be multiplied with an $\epsilon$ from the projection resulting 
in $\frac{1}{\epsilon}$ poles.
Therefore, in the $\oZ{2,1}{}$ mixing matrix all the 
blocks are non-trivial
\begin{equation}
\oZ{2,1}{}  = \begin{bmatrix} \oZ{2,1}{\Q\Q} & \oZ{2,1}{\Q\E} \\ \oZ{2,1}{\E\Q}& \oZ{2,1}{\E\E} \end{bmatrix} ~.
\end{equation}
It is precisely because $\oZ{2,1}{\E\Q}$ and $\oZ{1,0}{\E\Q}$ 
originate from $\frac{1}{\epsilon^2}$ and 
$\frac{1}{\epsilon}$ loop-integral divergences, respectively, 
that these constants are related by an analogue of the 
RG identity of eq.~\eqref{eq:RGidentity}, namely,
\begin{equation}
\oZ{2,1}{\E\Q} = \frac 12 \oZ{1,0}{\E\Q}\oZ{1,1}{\Q\Q} +  \frac 12 \oZ{1,1}{\E\E}\oZ{1,0}{\E\Q} - \frac{1}{2} \beta^{(1,0)}\oZ{1,0}{\E\Q}~.\label{eq:RGidentityEva}
\end{equation}
A relation of this form holds in any scheme, if we replace 
$\oZ{1,0}{\E\Q}$ with $(-1)\times$ the finite term $A^{(1)}_{\E\Q}$ 
in the one-loop correction to the vertex. 
In the scheme we are adopting, $\oZ{1,0}{\E\Q}$ is chosen 
to cancel $A^{(1)}_{\E\Q}$, and for 
this reason we can write the above identity
solely in terms of renormalization constants. 

We can now appreciate the
motivation for this choice of finite terms: by inspection of  
formula~\eqref{eq:twoloopgamma} for the two-loop anomalous dimension,
we see that eq.~\eqref{eq:RGidentityEva} implies that $\gamma^{(2,0)}_{\E\Q} = 0$~! 
Therefore, in this scheme the block structure of the one-loop
ADM (eq.~\eqref{eq:oneloopblock} persists also
at two-loop order, i.e.,
\begin{equation}
\gamma^{(2,0)}  = \begin{bmatrix}\gamma^{(2,0)}_{\Q\Q} & \gamma^{(2,0)}_{\Q\E}  \\ 0 & \gamma^{(2,0)}_{\E\E}  \end{bmatrix}~. \label{eq:twoloopblock}
\end{equation}
For applications to $d=4$ physics, 
this scheme has the advantage of enabling us to solve the RG flow 
without specifying the actual values of $C_{\E}$ when $d\to 4$ \cite{Herrlich:1994kh}.
Indeed, the finite subtraction in eq.~\eqref{eq:oneloopfinite} was first 
introduced for the calculation of the QCD NLO anomalous dimension 
of four-fermion operators in refs.~\cite{Buras:1989xd, Dugan:1990df}, 
and in a different but equivalent language in the context of $d=2$ 
Gross--Neveu/Thirring models in refs.~\cite{Bondi:1989nq,Bondi:1988fp}.
In the following, we shall refer to this scheme as the 
``{\it flavor scheme}''.

In $d=4-2\epsilon$ on the other hand, the one-loop finite renormalization
introduces an additional term linear in $\epsilon$ in the 
one-loop ADM, namely
\begin{align}
\gamma^{(1)} = \gamma^{(1,0)} + \epsilon \gamma^{(1,-1)}\,,
\quad\text{with}\quad
\gamma^{(1,-1)}  = 2 \Z^{(1,0)} = 2 \begin{bmatrix} 0 & 0 \\ \oZ{1,0}{\E\Q} & 0 \end{bmatrix}~.
\end{align}
As we discussed in the previous section, the term $\gamma^{(1,-1)}$ 
plays a role in cancelling the scheme dependence of the scaling 
dimension at the fixed point. 
Recall from eq.~\eqref{eq:gammastar} 
that $\gamma_2^*$ depends also on $\gamma^{(1,-1)}$, and it thus inherits a non-zero 
off-diagonal $\E\Q$ block.
Therefore, as far as scaling dimensions are concerned 
the simplified block structure of eq.~\eqref{eq:twoloopblock} 
in $\gamma^{(2,0)}$ is not particularly helpful because it does not persist 
in $\gamma_2^*$.
Had we, instead, adopted the pure $\MS$ scheme, i.e., $\Z^{(1,0)} = 0$, 
$\gamma^{(1,-1)}$ would be zero, but the two-loop ADM 
 $\gamma^{(2,0)}$ would itself have a non-zero $\E\Q$ block.

Summarizing, we have shown that in $d=4-2\epsilon$ the 
ADM at the fixed point 
has an invariant $\Q\Q$ block at 
order $\epsilon$, i.e., $(\gamma_1^*)_{\E\Q} = 0$, but the block 
is no longer invariant when we include also $\epsilon^2$ terms, i.e., 
$(\gamma_2^*)_{\E\Q} \neq 0$, neither in pure $\MS$ nor in 
the flavor scheme.
As such, the $\order{\epsilon^2}$ corrections of the scaling dimensions
cannot be computed solely from the $\Q\Q$ entries.
This is particularly problematic in cases with infinitely many 
evanescent operators, as in the example of four-fermion operators 
in eq.~\eqref{eq:evaexample2}. The computation of the eigenvalues
in this case is the topic of the next section.

\section{The evanescent tower \label{sec:EvaTow}}

In this section, we show how to obtain the NLO IR scaling 
dimensions of physical operators in the presence of mixing with 
an infinite tower of evanescent operators. 
For concreteness, we demonstrate the method for a specific example, 
which, however,
should make clear how to apply it more generally.

We consider the example of four-fermion operators in QED in 
$d=4-2\epsilon$, with $\Nf$ flavors of four-component Dirac 
fermions $\Psi^a$, $a = 1,\dots,\Nf$, namely
\begin{align}
\Q_1 & = T^{ac}_{bd} (\overline{\Psi}_a \gamma_\mu \Psi^b)(\overline{\Psi}_c \gamma_\mu \Psi^d)~,\\
\Q_3 & = T^{ac}_{bd} (\overline{\Psi}_a \Gamma^3_{\mu_1\mu_2\mu_3} \Psi^b)(\overline{\Psi}_c \Gamma^{3\,\mu_1\mu_2\mu_3} \Psi^d)~,
\end{align}
where the sum over repeated flavor indices is implicit. 
For the application of this example to  
the dynamics of QED in $d=3$, see ref.~\cite{DiPietroStamou2}. 
The tensor $T^{ac}_{bd} = T^{ca}_{db}$ specifies the flavor structure. 
In particular, we consider the ``flavor-nonsinglet'' 
case, for which $T^{ac}_{ad} = 0$ and $T^{ab}_{bd} = 0$, and the ``flavor-singlet''
case, for which $T^{ac}_{bd} = \delta^a_b \delta^c_d$. 
Since the interaction is flavor blind,
mixing does not spoil neither conditions on $T^{ac}_{bd}$.
 
The gauge coupling $\alpha = \frac{e^2}{16\pi^2}$ induces 
a mixing of the physical operators $(\Q_1,\Q_3)$ with the 
evanescent operators 
\begin{equation}
\E_n  = T^{ac}_{bd} (\overline{\Psi}_a \Gamma^{n}_{\mu_1\dots\mu_{n}} \Psi^b)(\overline{\Psi}_c \Gamma^{n\,\mu_1\dots\mu_{n}} \Psi^d) + a_n \epsilon \Q_1 + b_n \epsilon \Q_3~, \label{eq:defEvaBas}
\end{equation}
with $n$ running over all odd positive integers 
$\geq 5$. 
We have included terms proportional to $\epsilon$ 
with arbitrary coefficients $a_n,\,b_n$ as in ref.~\cite{Herrlich:1994kh}.
These terms reflect an intrinsic ambiguity in the definition of the 
evanescent operators. The final result for the scaling dimensions should not depend on these coefficients.
We shall use this as a check of our computation.
We do not include pieces of the form $\epsilon^2 \times$ a physical operator
because they have no effect in the two-loop computation presented here.
Since the expressions for the mixing matrices in this general 
basis are rather involved we set $a_n = b_n = 0$ in the rest of 
this section. 
We give the results in the more general basis 
in appendix~\ref{app:AnoDim}.
In the following we use pairs of odd integers $(n,m)$ 
as indices for matrices: the indices $1$ and $3$ refer to the 
physical operators $\Q_1$ and $\Q_3$, respectively, 
while indices $n\geq 5$ refer to the associated evanescent operators $\E_n$. 

\subsection{Flavor-nonsinglet operators}
Let us consider first the flavor-nonsinglet case, i.e. $T^{ac}_{ad} = 0 = T^{ab}_{bd}$.
First, we present the results in the flavor scheme and subsequently
perform  a change of scheme. 
The one- and two-loop ADM are given in appendix~\ref{app:AnoDim}. 
Furthermore, for QED we have 
that $\beta^{(1,0)} = - \frac 43 \Nf$ and $\beta^{(2,0)} = -4 \Nf$. 
Using eq.~\eqref{eq:gamma1star} we obtain that the $\order{\epsilon}$ 
anomalous dimension at the fixed point is
\begin{equation}
(\gamma_1^*)_{nm} = \frac{3}{2 \Nf}\times
\left\{
\begin{array}[]{cl}
n(n-1)(n-5)(n-6)&\text{for}\quad m=n-2\\
-2(n-1)(n-3)	&\text{for}\quad m=n\\
1 		&\text{for}\quad m=n+2\\
0               & \text{otherwise\,.}
\end{array}
\right.
\label{}
\end{equation}
All the $\E\Q$ entries in $\gamma_1^*$ vanish, in agreement 
with the argument of the previous section.
The result of the one-loop ADM for this 
nonsinglet case is also found in ref.~\cite{Dugan:1990df}.

At NLO in $\epsilon$, we obtain via eq.~\eqref{eq:gamma2star}
that the $\Q\Q$ block of the anomalous dimension at the fixed point is
\begin{equation}
(\gamma_2^*)_{\Q\Q} =
-\frac{1}{8 \Nf^2}
\begin{bmatrix}
729 		& 153 + 2 \Nf\\
324 + 792 \Nf	& -351 + 96\Nf
\end{bmatrix}  ~.
\end{equation}
This result derives from a two-loop computation of the corresponding 
renormalization constants.
For details on the computation we refer to
ref.~\cite{DiPietroStamou2}.
In the $\E\Q$ block there is a single non-vanishing entry 
in the finite renormalization $\oZ{1,0}{\E\Q}$, namely
\begin{equation}
\Z^{(1,0)}_{53} = - 40\,.
\end{equation}
It leads to a corresponding non-vanishing entry in the NLO 
ADM at the fixed point
\begin{equation}
(\gamma_2^*)_{53} =-\frac{60}{\Nf}\,, \label{eq:EQoffdiag}
\end{equation}
which, as we explained, hinders us from extracting the 
scaling dimensions of physical operators
solely from the $\Q\Q$ block.

To reduce the problem to a finite-dimensional one, we 
need to set the $\E\Q$ entries of the 
ADM to zero at NLO. 
This can be achieved either by a change of basis 
or equivalently by a change of scheme. We adopt 
the latter approach. We denote the finite renormalization 
in the new scheme by 
$\tilde{\Z}^{(1,0)}_{\E\Q} = \left(\tilde{\Z}^{(1,0)}_{n 1}, \,\tilde{\Z}^{(1,0)}_{n 3}\right)$. 
From the scheme shift in eq.~\eqref{eq:schgamma2star} 
we obtain the expression for the $\E\Q$ entries of 
$\gamma_2^*$ in the new scheme. 
Requiring
\begin{equation}
(\tilde{\gamma}_2^*)_{\E\Q} = 0~,
\end{equation} 
we find 
a recurrence relation
\begin{align}
  \tilde{\Z}^{(1,0)}_{(n + 2) 1}& 
  -2(n-1)(n-3) \tilde{\Z}^{(1,0)}_{n 1}
  + n(n-1)(n-5)(n-6)  \tilde{\Z}^{(1,0)}_{(n-2) 1} 
  - 36 \tilde{\Z}^{(1,0)}_{n 3} 
  \nonumber \\& =  1440 \,\delta_{n5}~, \label{eq:recrel1}\\
 \tilde{\Z}^{(1,0)}_{(n + 2) 3}& 
 -2(n-1)(n-3) \tilde{\Z}^{(1,0)}_{n 3}
 + n(n-1)(n-5)(n-6) \tilde{\Z}^{(1,0)}_{(n-2) 3} 
 -  \tilde{\Z}^{(1,0)}_{n 1} 
 \nonumber \\ & = \frac{160}{3}(12-\Nf) \,\delta_{n5} -  3360 \, \delta_{n 7}\,, \label{eq:recrel2}
\end{align}
which we need to solve to find the value of the new constants 
$(\tilde{\Z}^{(1,0)}_{n1},\tilde{\Z}^{(1,0)}_{n3})$. 
The index $n$ in the equations runs over the odd integers $\geq 5$. 
Since the recurrence relation is of second order, two boundary 
conditions are required. 
The first boundary condition is 
$\tilde{\Z}^{(1,0)}_{31} = \tilde{\Z}^{(1,0)}_{33} = 0 $, 
which means that we do not introduce any finite renormalization 
in the $\Q\Q$ block. As a second boundary condition, we  
require that $(\tilde{\Z}^{(1,0)}_{n1},\tilde{\Z}^{(1,0)}_{n3})$ 
do not grow too fast as $n\to \infty$, where ``too fast''
will be specified in a moment.

In the new scheme, the computation of the physical eigenvalues is reduced to the diagonalization of the invariant $\Q\Q$ block. The NLO $\Q\Q$ block reads
\begin{equation}
(\tilde{\gamma}_2^*)_{\Q\Q} = (\gamma_2^*)_{\Q\Q}  - \frac{9}{8 \Nf^2}\begin{bmatrix}
0		                &     0 \\
\tilde{\Z}^{(1,0)}_{51} & \tilde{\Z}^{(1,0)}_{53} + 40
\end{bmatrix}  ~.
\end{equation}
From eqs.~\eqref{eq:IRscaldimpert} and \eqref{eq:D1andD2} we then find the IR scaling dimension
of the $\Q$ operators to equal
\begin{equation}
(\Delta_{\IR})_i  = 2(d-1) + \epsilon (\Delta_1)_i + \epsilon^2 (\Delta_2)_i + \order{\epsilon^3}\,,
\end{equation}
with
\begin{align}
\label{eq:offD1D21}
(\Delta_1)_1 & 
= - \frac{9}{\Nf}\,,& 
(\Delta_2)_1 & 
= +\frac{3}{32 \Nf^2}\left(32 \Nf + 156 + \tilde{\Z}^{(1,0)}_{51} - 6 \tilde{\Z}^{(1,0)}_{53}\right)\,,& \\
\label{eq:offD1D22}
(\Delta_1)_2 & 
= + \frac{9}{\Nf}\,,& 
(\Delta_2)_2 & 
= -\frac{3}{32 \Nf^2}\left(160 \Nf +1140 + \tilde{\Z}^{(1,0)}_{51} + 6 \tilde{\Z}^{(1,0)}_{53}\right)\,.& 
\end{align}
By substituting the values 
of $(\tilde{\Z}^{(1,0)}_{51},\tilde{\Z}^{(1,0)}_{53})$ that solve the
 recurrence relation, we determine 
the value of the scaling dimension at NLO.

In practice, we use the following algorithm
to solve the recurrence relation: 
\begin{enumerate}
\item We truncate the recurrence relation by setting an
 upper cutoff $\ntr$ to the index $n$, i.e., 
we only consider the equations with $n < \ntr$;
\item We solve the resulting system of linear equations, 
treating $(\tilde{\Z}^{(1,0)}_{\ntr1},\,\tilde{\Z}^{(1,0)}_{\ntr3})$ 
as free parameters. The solution depends linearly on them.
Let us denote the solution for 
$(\tilde{\Z}^{(1,0)}_{51},\tilde{\Z}^{(1,0)}_{53})$ as
\begin{equation}
\label{eq:recsolution}
\begin{bmatrix} \tilde{\Z}^{(1,0)}_{51} \\ \tilde{\Z}^{(1,0)}_{53}
\end{bmatrix} = \begin{bmatrix} 
A_{51}(\ntr) \\ 
A_{53}(\ntr)  
\end{bmatrix} + B(\ntr)\cdot 
\begin{bmatrix} \tilde{\Z}^{(1,0)}_{\ntr1} \\ \tilde{\Z}^{(1,0)}_{\ntr3} 
\end{bmatrix}~,
\end{equation}
where $A_{51}(\ntr)$ and $A_{53}(\ntr)$ are constants that depend on the truncation
point $\ntr$ but not on $\tilde{\Z}^{(1,0)}_{\ntr1}$ or $\tilde{\Z}^{(1,0)}_{\ntr3}$, and
$B(\ntr)$ is an $\ntr$-dependent $2\times2$ matrix;
\item We impose the boundary condition
\begin{equation}
\lim_{\ntr\to \infty }B(\ntr)\cdot 
 \begin{bmatrix} 
   \tilde{\Z}^{(1,0)}_{\ntr1} \\ 
   \tilde{\Z}^{(1,0)}_{\ntr3} 
 \end{bmatrix} = 0~.\label{eq:bcinfinity}
\end{equation} 
This is the precise sense in which we require $\tilde{\Z}^{(1,0)}_{\ntr1}$ 
not to grow ``too fast''. It follows from this condition that 
\begin{equation}
\begin{bmatrix} \tilde{\Z}^{(1,0)}_{51} \\ \tilde{\Z}^{(1,0)}_{53} \end{bmatrix} 
=\lim_{\ntr\to \infty} 
\begin{bmatrix} 
A_{51}(\ntr) \\ 
A_{53}(\ntr) 
\end{bmatrix}\,.\label{eq:limittruncation}
\end{equation}
It is necessary for the consistency of the algorithm and in 
particular for the consistency of the boundary condition 
of eq.~\eqref{eq:bcinfinity}, that this limit exists and is finite.
Note that $B(\ntr)$ and $(\tilde{\Z}^{(1,0)}_{\ntr1},\,\tilde{\Z}^{(1,0)}_{\ntr3})$ 
depend on the normalization of the evanescent operators, 
but this normalization dependence drops in the product of eq.~\eqref{eq:bcinfinity}.
In what follows, we shall refer to  the $\ntr$ approximation,  
$(\Delta_2)_i[\ntr]$, as the value of $(\Delta_2)_i$ obtained substituting
 $(\tilde{\Z}^{(1,0)}_{51},\tilde{\Z}^{(1,0)}_{53})$ with $(A_{51}[\ntr],A_{53}[\ntr])$. 
 We verify the existence of the limit in eq.~\eqref{eq:limittruncation} by testing the convergence of $(\Delta_2)_i[\ntr]$.
\end{enumerate}
In a nutshell, this algorithm simply consists in truncating the 
infinite-dimensional matrix to a finite size, finding the 
first two eigenvalues for the truncated matrix, and then 
taking the limit in which the truncation is removed. 

\begin{figure}[t]
\begin{center}
\includegraphics{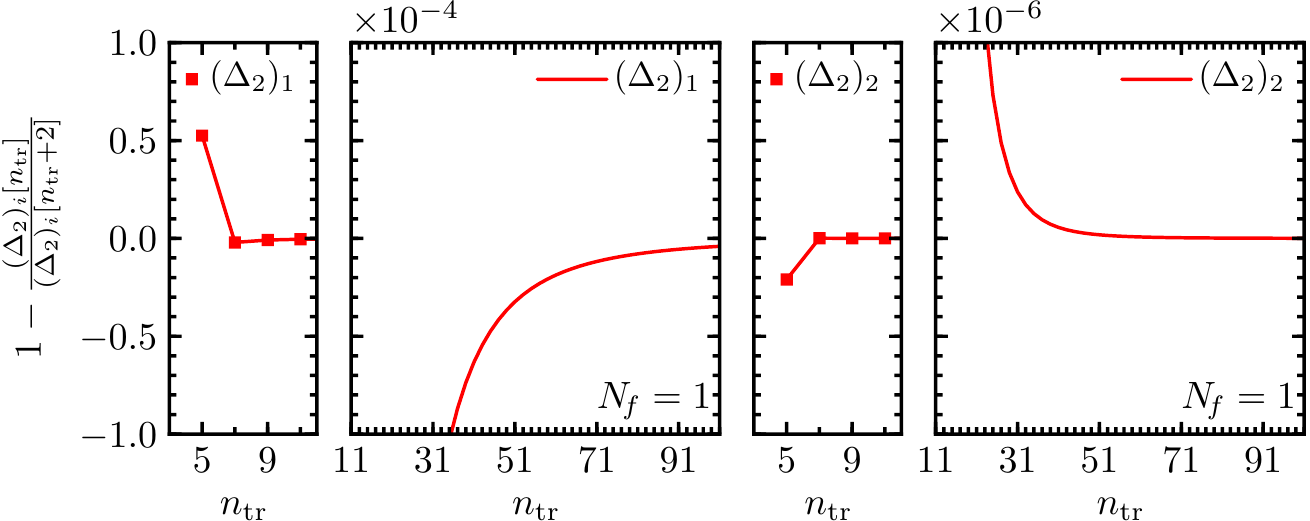}
\end{center}
\caption{
For the flavor-nonsinglet four-fermion operators we 
compute the  $\ntr$ and the $\ntr+2$ approximation to the two $(\Delta_2)_i$'s.
We plot the change between two neighbouring approximations, i.e., 
 $1-\frac{(\Delta_2)_i[\ntr]}{(\Delta_2)_i[\ntr+2]}$,
as function of the truncation point $\ntr$ for the case $\Nf=1$.
The left, right figure shows the truncation dependence of $(\Delta_2)_1$
and $(\Delta_2)_2$, respectively.
}
\label{fig:offdiagonalconvergence}
\end{figure}

We implemented this algorithm for different values of the parameter $\Nf$. 
Figure~\ref{fig:offdiagonalconvergence} shows how the NLO contribution to
the scaling dimension $(\Delta_2)_i$ relaxes as we increase the point of truncation.
To demonstrate this we plot the change in the approximation of $(\Delta_2)_i$ when the $\ntr$ is increased by 2, i.e,
$1-\tfrac{(\Delta_2)_i[\ntr]}{(\Delta_2)_i[\ntr+2]}$ , 
as a function of $\ntr$ for the case of $\Nf=1$. The behavior for larger $\Nf$ is analogous.
The plots show that as $\ntr$ increases the solution approaches 
a constant value, indicating that the limit in eq.~\eqref{eq:limittruncation} 
indeed exists.
In table~\ref{tab:offdiagonalD2}
we list the values of $(\Delta_2)_{1,2}$ for $\Nf=1,\dots,10$ 
for a truncation point so large that the significant digits displayed
are stable.
For comparison, we also show the LO values 
$(\Delta_1)_{1,2}$. 
Note, that with this choice of basis, i.e., $a_n=b_n=0$,
diagonalizing only the physical--physical block in the flavor scheme, i.e., not accounting for 
evanescent operators, amounts to a sizable 
numerical error. For instance, for $\Nf = 1$ we find that
$1-\frac{(\Delta_2)_i[\ntr=5]}{(\Delta_2)_i}=51\%, -21\%$ for $i=1,2$, respectively.

\begin{table}
\makebox[\textwidth][c]{
\begin{tabular}{lrrrrrrrrrr}
\multicolumn{1}{r}{$\boldsymbol{\Nf}$} 		& $1$ & $2$ & $3$ & $4$ & $5$ & $6$ & $7$ & $8$ & $9$ & $10$ \\
\hline\hline
$\boldsymbol{(\Delta_{1})_1}$	& $-9.00$ & $-4.50$ & $-3.00$ & $-2.25$ & $-1.80$ & $-1.50$ & $-1.29$ & $-1.12$ & $-1.00$ & $-0.900$\\
$\boldsymbol{(\Delta_{2})_1}$	& $35.6$ & $8.53$ & $3.63$ & $1.95$ & $1.19$ & $0.782$ & $0.544$ & $0.393$ & $0.292$ & $0.221$\\[0.5em]
$\boldsymbol{(\Delta_{1})_2}$	& $9.00$ & $4.50$ & $3.00$ & $2.25$ & $1.80$ & $1.50$ & $1.29$ & $1.12$ & $1.00$ & $0.900$\\
$\boldsymbol{(\Delta_{2})_2}$   & $-101$ & $-29.3$ & $-14.9$ & $-9.40$ & $-6.67$ & $-5.09$ & $-4.08$ & $-3.38$ & $-2.87$ & $-2.49$\\\hline
\end{tabular}}
\caption{
Three significant digits of the one-loop, $(\Delta_{1})_i$, and the two-loop, $(\Delta_{2})_i$, 
contributions to the scaling dimension of the flavor-nonsinglet operators for
various cases of $\Nf$.
To obtain the two-loop $(\Delta_{2})_i$ values we implemented the 
algorithm to include the effect of evanescent operators.
Higher truncation of the procedure does not affect the three significant digits displayed here.
}
\label{tab:offdiagonalD2}
\end{table}

In appendix~\ref{app:anbn} we also include the arbitrary coefficients
$a_n$ and $b_n$ of eq.~\eqref{eq:defEvaBas}. 
While the truncated solutions
depend linearly on $a_n$ and $b_n$, we show that the coefficients of the terms
proportional to $a_n$ and $b_n$ decrease to zero as we increase $\ntr$. 
This is an important check that the answer we obtain is indeed
a physical observable, independent of the choice of basis and renormalization scheme.

\subsection{Flavor-singlet operators}
We use the same approach to compute the NLO scaling dimensions for 
the flavor-singlet four-fermion operators for which $T^{ac}_{bd} = \delta^a_b\delta^c_d$, 
i.e.,
\begin{align}
\Q_1 & = (\overline{\Psi}_a \gamma_\mu \Psi^a)^2~,\\
\Q_3 & = (\overline{\Psi}_a \Gamma^3_{\mu_1\mu_2\mu_3} \Psi^a)^2\,.
\end{align}
Since the traces of $T^{ac}_{bd}$ are not zero, there are more 
diagrams contributing. As a result the ADM is not the same as
in the flavor-nonsinglet case, and there more non-zero entries 
of the mixing matrix compared to the flavor-nonsinglet case. 
In particular, while the flavor-nonsinglet case 
had a single non-zero $\E\Q$ entry at NLO (see eq.~\eqref{eq:EQoffdiag}), 
there are infinitely many non-zero entries in the flavor-singlet case. 
In addition to the $(5,3)$ entry of eq.~\eqref{eq:EQoffdiag},
we find that
\begin{align}
(\gamma_2^*)_{n 1} = \frac{24}{\Nf} (-1)^{\frac{n(n-1)}{2}}(n-2)(n-5)!~,\label{eq:diagEQ1}
\end{align}
where $n$ runs over all odd positive integers $\geq 5$. 
In terms of the general flavor tensor, 
this contribution is proportional to $T^{ab}_{bd}$, which
explains why it vanishes in the flavor-nonsinglet case. 
To compute~eq.~\eqref{eq:diagEQ1} we use the identity \cite{Kennedy:1981kp}
\begin{align}
\Gamma^n_{\mu_1\dots\mu_n}\gamma^\nu \Gamma^{n\,\mu_1\dots\mu_n} 
& \overset{\phantom{\epsilon\to0}}{=} (-1)^\frac{n(n-1)}{2}\frac{\Gamma(d)}{\Gamma(d-n+1)} (-1)^n (d-2n) \, \gamma^\nu \nonumber\\
& \overset{\epsilon\to 0}{=} -(-1)^\frac{n(n-1)}{2}24(n-2)(n-5)! \,\epsilon\,\gamma^\nu + \order{\epsilon^2}~.
\end{align}
We collect  the results for the ADM in appendix~\ref{app:AnoDim}.
The LO ADM at the fixed point then follows from them; it reads
\begin{equation}
(\gamma_1^*)_{nm} = \frac{3}{2 \Nf}\times
\left\{
\begin{array}[]{cl}
8 \delta_{n3} + n(n-1)(n-5)(n-6)&\text{for}\quad m=n-2\\
\frac 43 (2 \Nf +1)\delta_{n1}-2(n-1)(n-3)	&\text{for}\quad m=n\\
1 		&\text{for}\quad m=n+2\\
0               & \text{otherwise}
\end{array}
\right.
\label{}
\end{equation}
and the physical--physical block of the NLO ADM at the fixed point is
\begin{equation}
(\gamma_2^*)_{\Q\Q} =
-\frac{1}{24 \Nf^2}
\begin{bmatrix}
2383 + 224 \Nf 		& 375 + 18 \Nf\\
-1212 - 2568 \Nf	& -1485 - 360 \Nf
\end{bmatrix}~.
\end{equation}

\begin{figure}[t]
\begin{center}
\includegraphics{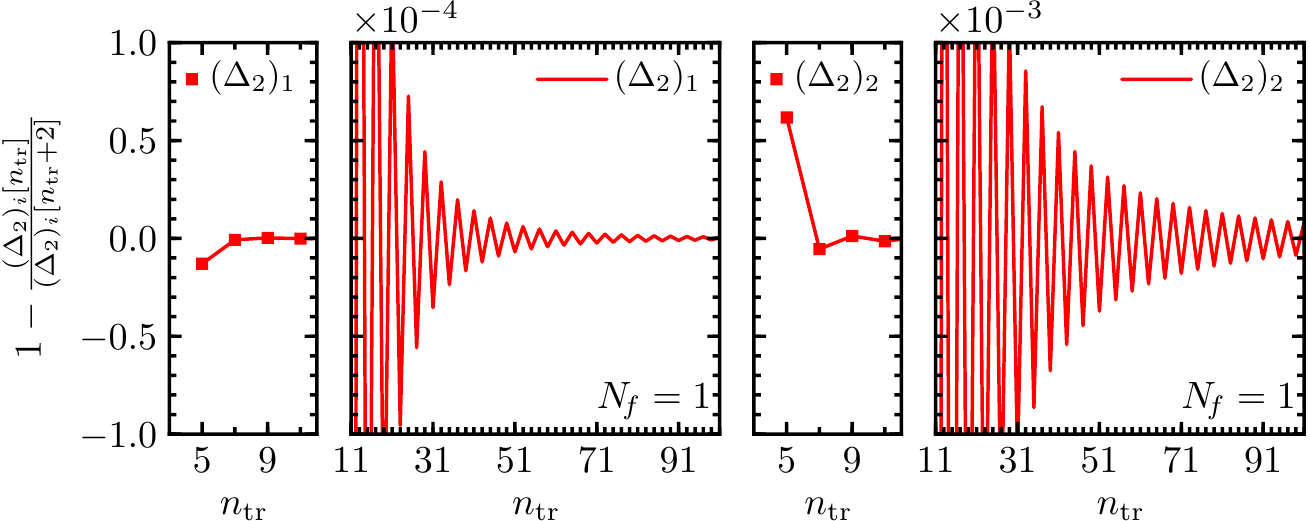}
\end{center}
\caption{
For the flavor-singlet four-fermion operators we 
compute the  $\ntr$ and the $\ntr+2$ approximation to the two $(\Delta_2)_i$'s.
We plot the change between two neighbouring approximations, i.e., 
 $1-\frac{(\Delta_2)_i[\ntr]}{(\Delta_2)_i[\ntr+2]}$,
as function of the truncation point $\ntr$ for the case $\Nf=1$.
The left, right figure shows the truncation dependence of $(\Delta_2)_1$
and $(\Delta_2)_2$, respectively.
}
\label{fig:diagonalconvergence}
\end{figure}

Analogously to the previous section we perform a change of scheme
and fix the finite renormalization constants by requiring 
that the $\E\Q$ entries of $\gamma_2^*$ vanish in the new scheme. 
This requirement defines  a recurrence relation analogous to the one of  
eqs.~\eqref{eq:recrel1} and \eqref{eq:recrel2}, which we solve with the same 
algorithm as above.
Note that the presence of infinitely many off-diagonal entries does not
change qualitatively the procedure. 
Also in this case we see that the solution
converges to a constant as we increase the size of the truncation, as shown in 
figure~\ref{fig:diagonalconvergence}, which is the analogue of 
figure~\ref{fig:offdiagonalconvergence} for the flavor-singlet case.
We list the values of the $\order{\epsilon^2}$ corrections, $(\Delta_2)_i$,
for $\Nf=1,\dots,10$ in table~\ref{tab:diagonalD2}.
For comparison we also list the LO values, $(\Delta_1)_i$, for the respective
$\Nf$ values.
Note that with this choice of basis, i.e., $a_n=b_n=0$,
diagonalizing only the physical--physical block,  
amounts for $\Nf = 1$ to a numerical error of  
$1-\frac{(\Delta_2)_i[\ntr=5]}{(\Delta_2)_i}=-14\%, 140\%$ for $i=1,2$,
respectively.
Also for this flavor-singlet case, we demonstrate in appendix~\ref{app:anbn} 
the independence of the scaling dimension on the 
parameters $a_n$ and $b_n$ from eq.~\eqref{eq:defEvaBas}.

\begin{table}
\makebox[\textwidth][c]{
\begin{tabular}{lrrrrrrrrrr}
\multicolumn{1}{r}{$\boldsymbol{\Nf}$} 		& $1$ & $2$ & $3$ & $4$ & $5$ & $6$ & $7$ & $8$ & $9$ & $10$ \\
\hline\hline
$\boldsymbol{(\Delta_{1})_1}$	& $-7.39$ & $-3.07$ & $-1.72$ & $-1.10$ & $-0.766$ & $-0.562$ & $-0.429$ & $-0.337$ & $-0.272$ & $-0.224$\\
$\boldsymbol{(\Delta_{2})_1}$	& $46.1$ & $14.1$ & $7.43$ & $4.84$ & $3.51$ & $2.73$ & $2.21$ & $1.86$ & $1.59$ & $1.39$\\[0.5em]
$\boldsymbol{(\Delta_{1})_2}$	& $13.4$ & $8.07$ & $6.39$ & $5.60$ & $5.17$ & $4.90$ & $4.71$ & $4.59$ & $4.49$ & $4.42$\\
$\boldsymbol{(\Delta_{2})_2}$   & $-84.0$ & $-23.5$ & $-11.6$ & $-7.12$ & $-4.94$ & $-3.70$ & $-2.91$ & $-2.37$ & $-1.99$ & $-1.70$\\\hline
\end{tabular}}
\caption{
Three significant digits of the one-loop, $(\Delta_{1})_i$, and the two-loop, $(\Delta_{2})_i$, 
contributions to the scaling dimension of the flavor-singlet operators for
various cases of $\Nf$.
To obtain the two-loop $(\Delta_{2})_i$ values we implemented the 
algorithm to include the effect of evanescent operators.
Higher truncation of the procedure does not affect the three significant digits displayed here.
}
\label{tab:diagonalD2}
\end{table}

\section{Conclusions and future directions}
We studied operator mixing  involving an infinite family of evanescent operators
in a $d$-dimensional theory, showing
how to extract the scaling dimensions at the fixed point 
beyond leading order in $\epsilon$.
At $\order{\epsilon^2}$, the scaling dimension is sensitive to the  one-loop mixing of the 
{\it whole tower} of operators. 
We demonstrated the independence of scaling dimensions
on the choice of basis and renormalization scheme.
We explicitly computed the $\order{\epsilon^2}$ corrections
in the example of four-fermion operators in QED.

In light of our findings, it would be interesting to revisit
the $\order{\epsilon^4}$ computation of the scaling dimension 
of the four-fermion interaction in the Gross--Neveu 
model from ref.~\cite{Gracey:2016mio}.
Since in this case the evanescent operators are first generated 
at three loops, we expect the whole evanescent tower to affect the 
$\order{\epsilon^4}$ term of the scaling dimension.

In the present work we only applied our method to extract the
first few eigenvalues of the ADM, whose eigenoperators
approach the physical operators for $\epsilon\to 0$.
The same procedure can also be applied to obtain additional 
eigenvalues. It would be interesting to study 
whether the additional eigenoperators also approach the physical operators
as $\epsilon \to 0$, or whether they are evanescent. 
The first case would mean that there exist multiple continuations of 
the physical operators to non-integer dimension and correspondingly multiple
functions that continue their scaling dimensions.
Another aspect that we have not addressed in this
work and that deserves further investigation is the loss
of unitarity of the $d$-dimensional CFT.
In analogy to refs.~\cite{Hogervorst:2014rta, Hogervorst:2015akt}, 
we expect that among the tower of evanescent operators
one may find states of negative norm, and operators of complex 
scaling dimensions.

\vspace{1em}
\noindent{\bf Acknowledgements:} 
we thank Joachim Brod, Martin Gorbahn, John Gracey, Zohar Komargodski, Giampiero Paffuti,
and David Stone for their interest and the many helpful discussions.
We are also indebted to the Weizmann Institute of Science, in which this research
began.
Research at Perimeter Institute is supported
by the Government of Canada through Industry Canada and by the Province of Ontario
through the Ministry of Research \& Innovation.

\appendix

\section{RG equations and scaling dimensions in $\boldsymbol{d = 4-2\epsilon}$\label{app:Resummation}}
In this appendix we review how to derive the expression for the scaling 
dimensions in terms of the ADM 
at the fixed point. This is usually done by solving the RG equation 
for the renormalized two-point function \cite{Mitter:1973ue, Brezin:1973jc, Brezin:1974eb}. 
The present derivation will emphasize the viewpoint that the renormalization 
constants are resumming the leading contributions for small $\epsilon$. We use this to take the IR limit of the ``bare'' two-point function.

Consider a set of operators $(\cO_0)_i$ 
that mix under the RG. We use the subscript ``$0$'' to distinguish 
them from the renormalized operators, whose correlators 
have a smooth $\epsilon \to 0$ limit. Recall that in 
$d = 4 - 2 \epsilon$ we have a dimensionful coupling 
$\alpha_0$ of dimension $2\epsilon$. 
When $|k|^{2\epsilon} \ll \alpha_0$ with $k$ the momentum of the operator
insertion, we can expand  the bare two-point function as
\begin{equation}
\langle (\cO_0)_i (-k) (\cO_0)_j(k) \rangle = |k|^{2 \Delta_{\UV} - d}\sum_{L=0}^{\infty} (\alpha_0 |k|^{-2 \epsilon})^L \sum_{M=-L}^{\infty}\rho_{0\,ij}^{(L,-M)}\epsilon^M~.\label{eq:bare}
\end{equation}
We are interested in the IR limit of this two-point function, namely the limit of 
large $\alpha_0 |k|^{-2 \epsilon}$. We keep $\epsilon\ll1$ and fixed. 

To constrain the two-point function we use 
input from the renormalized theory.
More precisely, we use the fact that there exist renormalized 
variables $\alpha$ and $\cO_j$ defined as 
\begin{align}
\alpha_0 & = Z_\alpha \alpha(\mu) \mu^{2 \epsilon} \label{eq:rencouplingApp} ~,\\
(\cO_0)_i & = \sum_j (\Z^{-1})_i^{~j} \cO_j  \label{eq:renoperatorApp} ~,
\end{align}
such that the renormalized two-point function, as a 
function of the renormalized coupling, has a smooth 
$\epsilon \to 0$ limit. The renormalized two-point 
function also has a perturbative expansion, i.e.,
\begin{equation}
\langle \cO_i (-k) \cO_j(k) \rangle  = |k|^{2 \Delta_{\UV}-d}\sum_{L=0}^{\infty}\alpha(\mu)^L \sum_{M=0}^L \rho^{(L,M)}_{ij} \log(k^2 / \mu^2)^M~, \label{eq:perturbative}
\end{equation}
where $\alpha(\mu)$ is the renormalized coupling, 
and $\mu$ the arbitrary renormalization scale. 
The negative powers of $\epsilon$ in eq.~\eqref{eq:bare} 
were chosen to match the powers of $\log(k^2/\mu^2)$ in eq.~\eqref{eq:perturbative}
as we take $\epsilon \to 0$.

From eqs.~\eqref{eq:expbeta} and \eqref{eq:expgamma} we have that
\begin{align}
- Z_\alpha \frac{\beta(\alpha, \epsilon)}{\alpha} = \frac{d \, Z_\alpha}{d \log \mu} &= (\partial_\alpha Z_\alpha)(-2 \epsilon \alpha + \beta(\alpha, \epsilon)) \,,\label{eq:Zalpha} \\
-\Z_i^{~k}\gamma(\alpha,\epsilon)_k^{~j} =\frac{d \,\Z_i^{~j}}{d \log \mu} &= (\partial_\alpha \Z_i^{~j})(-2 \epsilon \alpha + \beta(\alpha, \epsilon))\,.\label{eq:Zij}
\end{align}
The solution to these equations with boundary conditions 
$Z_\alpha(\alpha = 0) = 1$, $\Z_i^{~j}(\alpha = 0) = \mathbb{1}$ is
\begin{align}
Z_\alpha(\alpha, \epsilon) & = \exp\left(\int_0^{\alpha} d \alpha' \,\frac{\beta(\alpha', \epsilon)/\alpha'}{2 \epsilon \alpha'-\beta(\alpha', \epsilon)}\right)~, \label{eq:Zalphasol} \\
\Z(\alpha, \epsilon) & = \mathrm{\bar{P}exp}\left(\int_0^{\alpha} d \alpha' \,\frac{\gamma(\alpha', \epsilon)}{2 \epsilon \alpha'-\beta(\alpha', \epsilon)}\right) ~, \label{eq:Zmix}
\end{align}
up to the addition of functions of the variable $\epsilon$ alone, 
which are not important in what follows. $\mathrm{\bar{P}}$ 
denotes anti-path ordering. Note that 
the integral in the exponent of eq.~\eqref{eq:Zalphasol} is 
finite near the lower end $\alpha = 0$, because in this region
$\beta(\alpha, \epsilon) = - 2 \beta^{(1,0)} \alpha^2 + \order{\alpha^3}$.

The dimensionless parameter that becomes large in the IR limit is 
\begin{equation}
\alpha_0 |k|^{-2 \epsilon} = Z_\alpha \alpha(\mu)\left(\tfrac{\mu}{|k|}\right)^{2 \epsilon}~.
\end{equation}
Therefore, taking this IR limit while maintaining $\mu = |k|$, implies
that on the right-hand side  $Z_\alpha\alpha(\mu)$ must become large. 
As $\alpha$ grows continuously on the positive real axis, starting from 
the UV value $\alpha = 0$, the constant $Z_\alpha$ can become large 
if $\alpha$ approaches a pole of the integrand of eq.~\eqref{eq:Zalphasol}, i.e., 
a non-trivial fixed point $\alpha^*$. Close to the solution we expand
$-2\epsilon \alpha + \beta(\alpha, \epsilon) \sim C(\alpha - \alpha^*)$. 
Substituting this in the integral, 
we find the leading behavior of $\alpha_0 |k|^{-2 \epsilon}$ as $\alpha$ approaches the
fixed point
\begin{equation}
\alpha_0 |k|^{-2 \epsilon} \sim \alpha^* |\alpha - \alpha^*|^{-\frac{2\epsilon}{C}}~. \label{eq:IRparameter}
\end{equation}
Given that $\epsilon \ll 1$ we can see 
perturbatively that the fixed point exists when $\beta^{(1,0)}< 0$ and that $C > 0$. 
The latter inequality ensures that indeed $\alpha_0 |k|^{-2 \epsilon}$ grows as we
approach the fixed point.

Similarly to $Z_\alpha$, eq.~\eqref{eq:Zmix} implies that 
as $\alpha\to \alpha^*$ the leading behavior of $\Z_i^{~j}$ is
\begin{equation}
\Z_i^{~j} \sim |\alpha - \alpha^*|^{-\tfrac{\gamma(\alpha^*, \epsilon)}{C}}~.
\end{equation}
Using eq.~\eqref{eq:IRparameter}, we find that in the IR limit $\Z_i^{~j}$ 
becomes a power-law in $|k|$, namely
\begin{equation}
\Z_i^{~j} \sim \left( \frac{|k|}{\Lambda} \right)^{- \gamma(\alpha^*, \epsilon)}~,
\label{eq:crossover}
\end{equation}
where we introduced the crossover scale $\Lambda$, whose 
leading behavior as a function of $\epsilon$ for $\epsilon \ll 1$ is
\begin{equation}
\Lambda \sim \left(-\beta^{(1,0)}\frac{\alpha_0}{\epsilon}\right)^{\frac{1}{2\epsilon}}~.\label{eq:crossscale}
\end{equation}
Recalling that 
\begin{equation}
\langle (\cO_0)_i (-k) (\cO_0)_j(k) \rangle = (\Z^{-1})_i^{~k} (\Z^{-1})_j^{~l} \langle \cO_k (-k) \cO_l(k) \rangle~,
\end{equation}
and using eqs.~\eqref{eq:crossover} and \eqref{eq:perturbative} 
for $\mu = |k|$ we see that in the IR limit a new scaling behavior emerges, i.e.,
\begin{equation}
\langle (\cO_0)_i (-k) (\cO_0)_j(k) \rangle \underset{|k| \ll \Lambda}\sim 
 |k| ^{2(\Delta_\UV+\gamma(\alpha^*, \epsilon))-d}\,,
\label{eq:IRlimtwopt}
\end{equation}
corresponding to the IR scaling dimension
$\Delta_{\IR} = \Delta_{\UV} + \gamma(\alpha^*, \epsilon)$. 
We also see that more precisely the crossover to the IR scaling 
happens when $|k| \sim \Lambda$, with $\Lambda$ given in 
eq.~\eqref{eq:crossscale}. 
As observed in Ref.~\cite{DiPietro:2015taa}, 
$\Lambda$ is exponentially enhanced for $\epsilon \ll 1$ 
compared to the naive crossover scale $\alpha_0^{\frac{1}{2\epsilon}}$.

\section{Beta functions and anomalous dimensions\label{app:BetaAD}}
In this appendix we collect: 
\begin{enumerate}
\item[\it i)]  the one- and two-loop formulas for the beta function and 
ADM from eqs.~\eqref{eq:expbeta} and \eqref{eq:expgamma}, respectively.
In terms of the renormalization-constant expansions from  
eqs.~\eqref{eq:expconst1} and \eqref{eq:expconst2} they read
\begin{align}
\beta^{(1,0)}  & = - Z_\alpha^{(1,1)}~,\qquad
\beta^{(1,-1)} = -  Z_\alpha^{(1,0)}~,\nonumber\\
\beta^{(2,0)}  & = - 2 Z_\alpha^{(2,1)} + 4 Z_\alpha^{(1,0)} Z_\alpha^{(1,1)}~,\label{eq:twoloopbeta}\\[1em]
\gamma^{(1,0)} & = 2 \oZ{1,1}{}~,\qquad\gamma^{(1,-1)}=2 \oZ{1,0}{}~,\nonumber\\		  
\gamma^{(2,0)} & = 
      4 \oZ{2,1}{} - 2\oZ{1,1}{} \oZ{1,0}{}- 2\oZ{1,0}{}\oZ{1,1}{} 
     + 2\beta^{(1,0)} \oZ{1,0}{}
     +2  \beta^{(1,-1)}\oZ{1,1}{}~.\label{eq:twoloopgamma}
\end{align}
\item[\it ii)] the relations between the one- and two-loop 
	       beta function and ADM in two different, mass-independent
	       schemes distinguished by the superscript ``$\widetilde{\phantom{a}}$''.
	       Substituting the expansion of 
	       eqs.~\eqref{eq:expbeta} and \eqref{eq:oepsbeta} in eq.~\eqref{eq:chainbeta} 
	we find that
\begin{align}
\tilde{\beta}^{(1,0) }  &= \beta^{(1,0)}~,& 
\tilde{\beta}^{(1,-1)}
&=\beta^{(1,-1)}  + Z^{(1,0)} -\tilde{Z}^{(1,0)}~,&\nonumber \\
\tilde{\beta}^{(2,0) }  &= \beta^{(2,0)}~,&
\tilde{\beta}^{(2,-1)} &= \beta^{(2,-1)} +2 \left(
            Z^{(2,0)} - \tilde{Z}^{(2,0)}  
            - (Z^{(1,0)} )^2  + (\tilde{Z}^{(1,0)})^2 \right)~.&
\label{eq:pertchangebeta}
\end{align}
Similarly, the expansion of eqs.~\eqref{eq:expgamma} and \eqref{eq:oepsgamma} 
in eq.~\eqref{eq:schgamma} leads to
\begin{align}
\tilde{\gamma}^{(1,0) }&=\gamma^{(1,0)}~,\qquad\quad
\tilde{\gamma}^{(1,-1)} =\gamma^{(1,-1)}
-2\left(\oZ{1,0}{}-\tZ{1,0}{}\right)\\[0.5em]
\tilde{\gamma}^{(2,0)} &=\gamma^{(2,0)}+ 
     \left[\gamma^{(1,0)}, \oZ{1,0}{}- \tZ{1,0}{}\right] 
  - 2 \beta^{(1,0)} \left(\oZ{1,0}{}- \tZ{1,0}{}\right)~.
\label{eq:pertchangegamma}  
\end{align}
\end{enumerate}

\section{Anomalous dimensions of four-fermion operators in QED\label{app:AnoDim}}
\subsection{Flavor-nonsinglet operators}
The results for the one-loop anomalous dimension, and the one-loop 
finite renormalization in the flavor scheme can be found 
in ref.~\cite{Dugan:1990df}. Including also the dependence on $a_n,\,b_n$ 
the result is
\begin{align}
\gamma^{(1,0)}_{n m} &=
\left\{
\begin{array}[]{cl}
 2n(n-1)(n-5)(n-6)&\text{for}\quad m=n-2\\
-4(n-1)(n-3)     &\text{for}\quad m=n\\
2               &\text{for}\quad m=n+2\\
0               &\text{otherwise\,,}\\
\end{array}
\right.\\
\gamma^{(1,-1)}_{n m} &=
\left\{
\begin{array}[]{ll}
        - 2  n(n-1)(n-5)(n-6) a_{n-2}\\
        \quad+ 4 (n-1)(n-3) a_{n}- 2 a_{n+2} + 72 b_n
        &\text{for}~m=1,\,n\ge5\\[1em]
        - 80 \delta_{n5}\\
        \quad-  2n(n-1)(n-5)(n-6)  b_{n-2}\\
        \qquad+ 4(n-1)(n-3) b_{n}- 2b_{n+2}+ 2a_n 
        &\text{for}~m=3,\,n\ge5\\[1em]
\hspace*{7em}0 &\text{otherwise\,.}
\end{array}
\right.
\label{eq:gamma1min1off}
\end{align}
We computed also the $(\Q_1,\Q_3)$ entries of the two-loop anomalous 
dimension in the same scheme, finding
\begin{equation}
\begin{split}
\gamma^{(2,0)}_{\Q\Q} &=
\left[
\begin{array}{cc}
          - 162    & -28-\frac{4}{9}\Nf \\
            144 -176 \Nf & 78- \frac{64}{3}\Nf
\end{array}
\right]+\\
&+ a_5 \left[\begin{array}{cc}
        -2 & 0 \\
        -\frac{8}{3} \Nf & 2 \\
        \end{array}\right]
 + b_5 \left[\begin{array}{cc}
        0 & -2 \\
        72 & -\frac{8}{3} \Nf \\
        \end{array}\right]\,.
\end{split}
\label{eq:gamma20qqoff}
\end{equation}
Moreover, as explained in section \ref{sec:EvaOpe}, in this scheme 
$\gamma^{(1,0)}_{\E\Q}= \gamma^{(2,0)}_{\E\Q}= 0$.

Using eqs.~\eqref{eq:gamma1star} and \eqref{eq:gamma2star} we obtain the ADM at the
fixed point
\begin{align}
(\gamma_1^*)_{nm} &= \frac{3}{2 \Nf}\times
\left\{
\begin{array}[]{cl}
n(n-1)(n-5)(n-6)&\text{for}\quad m=n-2\\
-2(n-1)(n-3)    &\text{for}\quad m=n\\
1               &\text{for}\quad m=n+2\\
0               & \text{otherwise\,.}
\end{array}
\right.\\[2em]
(\gamma_2^*)_{nm} &=
\left\{
\begin{array}[]{ll}
	-\frac{1}{8 \Nf^2}
	\begin{bmatrix}
	729             & 153 + 2 \Nf\\
	324 + 792 \Nf   & -351 + 96\Nf
	\end{bmatrix}   & \\
	+\frac{3}{8\Nf^2}a_5
	\begin{bmatrix}
	-3 & 0\\
	-4\Nf & 3
	\end{bmatrix}
	+\frac{3}{8\Nf^2}b_5
	\begin{bmatrix}
	0 & -3\\
	108 & -4\Nf
	\end{bmatrix}
	&\text{for}\quad n,m=1,3\,,\\[2em]
	\frac{3}{2 \Nf}  \left( -n(n-1)(n-5)(n-6) a_{n-2} \right. \\
	\left.\qquad + 2 (n-1)(n-3)a_{n} -  a_{n+2} + 36 b_n \right)
	&\text{for}~m=1,\,n\ge5~,\\[1em]
	- \frac{60}{\Nf} \delta_{n5}\\
	\quad+ \frac{3}{2 \Nf}\left( - n(n-1)(n-5)(n-6)  b_{n-2} \right.\\
	\qquad\left. + 2(n-1)(n-3) b_{n}- b_{n+2}+ a_n \right)
	&\text{for}~m=3,\,n\ge5~,\\[1em]
	\hspace*{5em}\text{not required}  & \text{otherwise\,.}
\end{array}
\right.\label{eq:gamma2starabnsing}
\end{align}

\subsection{Flavor-singlet operators}

The one-loop anomalous dimension in the physical sector
can be found in refs.~\cite{Beneke:1995qq, DiPietro:2015taa}.
The one-loop $\E\Q$ finite terms and the two-loop $\Q\Q$ ADM 
in the flavor scheme are computed in 
ref.~\cite{DiPietroStamou2}. The results are
\begin{align}
\gamma^{(1,0)}_{n m} &=
\left\{
\begin{array}[]{cl}
16\delta_{n3}+ 2n(n-1)(n-5)(n-6)&\text{for}\quad m=n-2\\
\frac{8}{3}(2\Nf+1)\delta_{n1} -4(n-1)(n-3)	&\text{for}\quad m=n\\
2 		&\text{for}\quad m=n+2\\
0 		&\text{otherwise\,,}\\
\end{array}
\right.\\
\gamma^{(1,-1)}_{nm} &=
\left\{
\begin{array}[]{ll}
	 32 (-1)^{\frac{n(n-1)}{2}} (n-2)(n-5)!\\
	\quad- 2  n(n-1)(n-5)(n-6) a_{n-2}\\
	\qquad+ \left(\frac{8}{3}(2\Nf+1)+ 4 (n-1)(n-3)\right) a_{n}\\
	\qquad\quad- 2 a_{n+2} + 88 b_n
	&\text{for}~m=1,\,n\ge5\\[1em]
	- 80 \delta_{n5}\\
	\quad-  2n(n-1)(n-5)(n-6)  b_{n-2}\\
	\qquad+ 4(n-1)(n-3) b_{n}- 2b_{n+2}+ 2a_n 
	&\text{for}~m=3,\,n\ge5\\[1em]
\hspace*{7em}0 &\text{otherwise\,,}
\end{array}
\right.  \label{eq:gamma1min1}
\end{align}
and
\begin{equation}
\begin{split}
\gamma^{(2,0)}_{\Q\Q} &=
\left[
\begin{array}{cc}
          -  \frac{2}{27} (2275+8 \Nf)    & -\frac{4}{9}  (49+ 3 \Nf) \\
             \frac{16}{9} (199 + 107 \Nf) & 110+ \frac{80 \Nf}{3}
\end{array}
\right]+\\
&+ a_5 \left[\begin{array}{cc}
 	-2 & 0 \\
 	\frac{8}{3} (1+\Nf) & 2 \\
	\end{array}\right]
 + b_5 \left[\begin{array}{cc}
 	0 & -2 \\
 	88 & -\frac{8}{3} \Nf \\
	\end{array}\right]~.
\end{split} \label{eq:gamma20qq}
\end{equation}
Also in this case $\gamma^{(1,0)}_{\E\Q}= \gamma^{(2,0)}_{\E\Q}= 0$ 
in agreement with the general results of section \ref{sec:EvaOpe}.

Using eqs.~\eqref{eq:gamma1star} and \eqref{eq:gamma2star} we obtain the ADM at the
fixed point
\begin{align}
(\gamma_1^*)_{nm} &= \frac{3}{2 \Nf}\times
\left\{
\begin{array}[]{cl}
8 \delta_{n3} + n(n-1)(n-5)(n-6)&\text{for}\quad m=n-2\\
\frac 43 (2 \Nf +1)\delta_{n1}-2(n-1)(n-3)	&\text{for}\quad m=n\\
1 		&\text{for}\quad m=n+2~\\
0               & \text{otherwise}~,
\end{array}
\right.\label{}\\[1em]
(\gamma_2^*)_{nm} &=
\left\{
\begin{array}[]{ll}
	-\frac{1}{24 \Nf^2}
	\begin{bmatrix}
	2383 + 224 \Nf 		& 375 + 18 \Nf\\
	-1212 - 2568 \Nf	& -1485 - 360 \Nf
	\end{bmatrix}&\\
	+\frac{3}{8\Nf^2}
	a_5
	\begin{bmatrix}
	-3 & 0\\
	4\Nf+4 &3
	\end{bmatrix}
	+\frac{3}{8\Nf^2}
	b_5
	\begin{bmatrix}
	0 & -3\\
	132 & -4\Nf
	\end{bmatrix}
	&\text{for}~n,m=1,3~,\\[2em]
	\frac{24}{\Nf} (-1)^{\frac{n(n-1)}{2}} (n-2)(n-5)!\\
	\quad + \frac{3}{2 \Nf}  \left( -n(n-1)(n-5)(n-6) a_{n-2} \right. \\
	\qquad +\left( \frac{4}{3}(2\Nf+1)+ 2 (n-1)(n-3)\right) a_{n}\\
	\qquad\quad \left. -  a_{n+2} + 44 b_n \right)
	&\text{for}~m=1,\,n\ge5~,\\[1em]
	- \frac{60}{\Nf} \delta_{n5}\\
	\quad+ \frac{3}{2 \Nf}\left( - n(n-1)(n-5)(n-6)  b_{n-2} \right.\\
	\qquad\left. + 2(n-1)(n-3) b_{n}- b_{n+2}+ a_n \right)
	&\text{for}~m=3,\,n\ge5~,\\[1em]
\hspace*{7em}\text{not required} &\text{otherwise\,.}
\end{array}
\right. \label{eq:gamma2starabsing}
\end{align}

\section{\texorpdfstring{%
$\boldsymbol{a_n}$ and $\boldsymbol{b_n}$ independence \label{app:anbn}}{%
a_n and b_n independence}}

\begin{figure}[t]
\begin{center}
\includegraphics{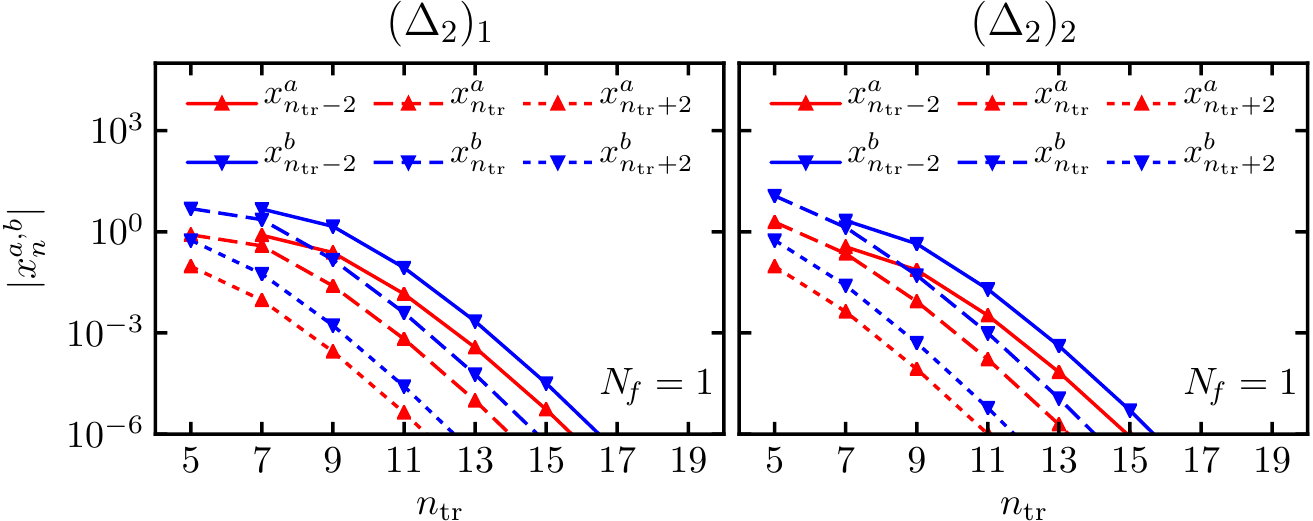}
\end{center}
\caption{
For the case of flavor-nonsinglet operators, we plot the coefficient of the basis-dependent parameters $a_n$ and $b_n$ in the truncated result for the NLO scaling dimension, as a function of the truncation number
$\ntr$, for $\Nf=1$. We see that as we increase the number of evanescent operators included,
the dependence drops from the observable.
See eq.~\eqref{eq:D2generic} for more details.
}
\label{fig:offdiagonalanbnconvergence}
\end{figure}
In appendix \ref{app:AnoDim} we collected the results for the ADM
for both the flavor-singlet and flavor-nonsinglet operators
computed in the generic basis of eq.~\eqref{eq:defEvaBas}.
We observe that both the $\E\Q$ finite one-loop mixing 
and the two-loop $\Q\Q$ mixing 
depends linearly on the coefficients $a_n$ and $b_n$.
As a result, in this generic basis,  the entries of $\gamma_2^*$, which 
we use to compute the NLO scaling dimension, also depend on $a_n$ and $b_n$.
However, the parameters $a_n$ and $b_n$ are just 
a parametrization of our freedom to choose the basis of operators.
Therefore, the observables, i.e., the scaling dimensions,
cannot depend on them.
In this appendix, we demonstrate using the algorithm from section~\ref{sec:EvaTow} how this unphysical 
dependence indeed cancels in the observables. 
This has to be contrasted with the wrong procedure of naively diagonalizing the $\Q\Q$ block of the two-loop ADM
in the flavor scheme, 
which would lead to eigenvalues that depend on $a_n$ and $b_n$.

To this end, we first generalize eqs.~\eqref{eq:offD1D21} and \eqref{eq:offD1D22}
to the case in which the basis includes $a_n$ and $b_n$.
The generalizations for the two considered cases 
follow:\\[0.5em]
{\it Flavor-nonsinglet case:}
\begin{align}
(\Delta_2)_1 & = 
\frac{3}{32 \Nf^2}\biggl(
156+32 \Nf
-\left(10-\frac{4}{3}\Nf\right) a_5
+a_7
+(60-8 \Nf)b_5
-6 b_7	\nonumber\\&\hspace{22em}
+ \tilde{\Z}^{(1,0)}_{51} 
- 6 \tilde{\Z}^{(1,0)}_{53}\biggr) ~, \\
(\Delta_2)_2 & = 
\frac{3}{32 \Nf^2}\biggl(
 -1140-160 \Nf 
+\left(22-\frac{4}{3}\Nf\right)a_5
-a_7
+(132-8\Nf)b_5
-6 b_7	\nonumber\\&\hspace{22em}
- \tilde{\Z}^{(1,0)}_{51} - 6 \tilde{\Z}^{(1,0)}_{53}\biggr)\,. 
\end{align}
{\it Flavor-singlet case:}
\begin{align}
(\Delta_2)_1  = 
\frac{1}{192 \Nf^2 \kappa}\biggl(
+ &47552 - 7912\kappa + (6992 + 544\kappa) \Nf + 2336 \Nf^2 \nonumber\\
- &  (1350 - 108\kappa) a_5\nonumber\\
+ & 81 a_7\nonumber\\
-& (2700 - 1728\kappa - (1656 - 144\kappa) \Nf + 144 \Nf^2) b_5 \nonumber\\
-& (54 + 108\kappa + 108 \Nf) b_7\nonumber\\
+ & 81 \tilde{\Z}^{(1,0)}_{51} 
-(54 + 108\kappa + 108 \Nf)\tilde{\Z}^{(1,0)}_{53} \biggr)\,,\\
(\Delta_2)_2  = 
\frac{1}{192 \Nf^2 \kappa}\biggl(
- & 47552 - 7912\kappa + (-6992 + 544\kappa) \Nf - 2336 \Nf^2 \nonumber\\
+ &(1350 + 108\kappa) a_5\nonumber\\
- & 81 a_7 \nonumber\\
+ & (2700 + 1728\kappa - (1656 + 144\kappa) \Nf + 144 \Nf^2) b_5\nonumber\\
+ & (54 - 108\kappa + 108 \Nf) b_7\nonumber\\
- &81 \tilde{\Z}^{(1,0)}_{51} 
+ (54 - 108\kappa + 108 \Nf) \tilde{\Z}^{(1,0)}_{53} \biggr)\,,
\end{align}
where we introduced 
\begin{equation}
\kappa\equiv\sqrt{\Nf^2+\Nf+25}\,.
\label{}
\end{equation}

We see that in both cases $(\Delta_2)_i$ depend linearly on $a_n$, $b_n$ with $n = 5,7$ and on 
the finite renormalization constants $\tilde{\Z}^{(1,0)}_{51}$ and $\tilde{\Z}^{(1,0)}_{53}$. 
By setting the $\E\Q$ entries of $\tilde{\gamma}_2^*$ in eqs.~\eqref{eq:gamma2starabnsing} and 
\eqref{eq:gamma2starabsing} to zero, we obtain a generalization of the recurrence relation 
in eqs.~\eqref{eq:recrel1} and \eqref{eq:recrel2} for $\tilde{\Z}^{(1,0)}$, 
with additional terms linear in $a_n$, $b_n$. 
After solving the truncated system of equations with $n < \ntr$ and 
substituting back the solution for $\tilde{\Z}^{(1,0)}_{51}$ and $\tilde{\Z}^{(1,0)}_{53}$
in terms of $\tilde{\Z}^{(1,0)}_{\ntr1}$ and $\tilde{\Z}^{(1,0)}_{\ntr3}$, 
we find that the dependence on $a_n$ and $b_n$ for $n<\ntr-2$ cancels.
The $\ntr$ approximation to the observable, $\Delta_2[\ntr]$, defined by 
\begin{equation}
\Delta_2 = 
\Delta_2[\ntr] 
+ x^\Z_1[\ntr]\, \tilde{\Z}^{(1,0)}_{\ntr 1} + x^\Z_3[\ntr]\, \tilde{\Z}^{(1,0)}_{\ntr 3}~,
\end{equation}
depends solely on higher $a_n$'s and
$b_n$'s, namely
\begin{align}
\begin{split}
\Delta_2[\ntr] = \Delta_2[\ntr]\biggl\vert_{a_n=b_n=0}+ &x^a_{\ntr-2}[\ntr]\, a_{\ntr-2} + x^a_{\ntr}[\ntr]\, a_{\ntr} + x^a_{\ntr+2}[\ntr]\, a_{\ntr+2} \\
+ &x^b_{\ntr-2}[\ntr]\, b_{\ntr-2} + x^b_{\ntr}[\ntr]\, b_{\ntr} + x^b_{\ntr+2}[\ntr]\, b_{\ntr+2} \,.
\end{split}
\label{eq:D2generic}
\end{align}

\begin{figure}[t]
\begin{center}
\includegraphics{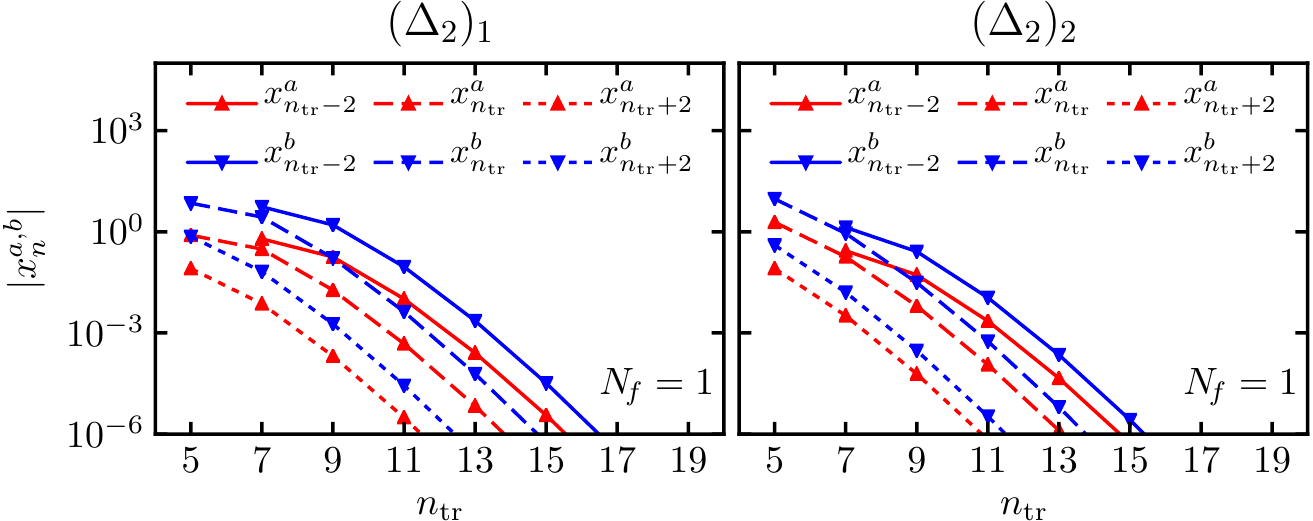}
\end{center}
\caption{
For the case of flavor-singlet operators, we plot the coefficient of the basis-dependent parameters $a_n$ and $b_n$ in the truncated result for the NLO scaling dimension, as a function of the truncation number
$\ntr$, for $\Nf=1$. We see that as we increase the number of evanescent operators included,
the dependence drops from the observable.
See eq.~\eqref{eq:D2generic} for more details.
}
\label{fig:diagonalanbnconvergence}
\end{figure}
Here, we have suppressed for brevity the subscript $i$ labelling the eigenvalue. 
Therefore, as a final check of the basis independence 
of the observable  we show that the coefficients of the terms proportional to 
$a_{\ntr-2}$, $a_{\ntr}$, $a_{\ntr+2}$, $b_{\ntr-2}$, $b_{\ntr}$, and $b_{\ntr+2}$
all relax to zero as $\ntr\to\infty$, i.e., that
\begin{align}
&\lim_{\ntr\to\infty} x^{a,b}_{\ntr-2}[\ntr] = 0\,,&
&\lim_{\ntr\to\infty} x^{a,b}_{\ntr}[\ntr] = 0\,,&
&\lim_{\ntr\to\infty} x^{a,b}_{\ntr+2}[\ntr] = 0\,.&
\end{align}
We demonstrate this in 
figure~\ref{fig:offdiagonalanbnconvergence} and 
figure~\ref{fig:diagonalanbnconvergence} for the flavor-nonsinglet and 
flavor-singlet case, respectively, in which we plot the absolute value of these 
coefficients as a function of $\ntr$
for the case $\Nf=1$.
In both cases, we observe that even after a few steps of the 
algorithm the coefficients have already relaxed to small values. The behavior for larger values of $\Nf$ is analogous.

\addcontentsline{toc}{section}{References}
\bibliographystyle{JHEP}
\bibliography{references}
\end{document}